\documentclass[fleqn,usenatbib]{mnras}

\usepackage{newtxtext,newtxmath}
\usepackage[T1]{fontenc}

\DeclareRobustCommand{\VAN}[3]{#2}
\let\VANthebibliography\thebibliography
\def\thebibliography{\DeclareRobustCommand{\VAN}[3]{##3}\VANthebibliography}

\usepackage{graphicx}	% Including figure files
\usepackage{amsmath}	% Advanced maths commands
\usepackage{verbatim} %才能多行注释
\usepackage{float}
\usepackage{stfloats}
\usepackage{multirow} %表格一个占多行
\usepackage{multicol} %表格一个占多列
\usepackage{supertabular}%表格跨页
\usepackage{CJKutf8}%中文

\title[IFS Study Of Thin \& Thick Discs In IC 2531]{An Integral Field Spectroscopy Study Of Thin And Thick Discs In IC 2531}

\author[Fan Xu et al.]{
Fan Xu (\begin{CJK*}{UTF8}{gbsn}徐钒\end{CJK*}),$^{1}$\thanks{E-mail: u7369768@alumni.anu.edu.au}
K. C. Freeman,$^{1}$\thanks{E-mail: Kenneth.Freeman@anu.edu.au}
Henry Zovaro$^{1}$
\\
$^{1}$Research School of Astronomy \& Astrophysics, Mount Stromlo Observatory, Cotter Road, Weston Creek, ACT 2611, Australia\\
}

\date{Accepted XXX. Received YYY; in original form ZZZ}

\pubyear{2015}

\begin{document}
\label{firstpage}
\pagerange{\pageref{firstpage}--\pageref{lastpage}}
\maketitle

\begin{abstract}
Thin and thick discs are prominent components in the Milky Way and other spiral galaxies, but their formation histories are not yet understood, which is particularly true for thick discs. This is partially due to the fact that we lack sufficient understanding of thin and thick discs in other galaxies to determine how common features of the Milky Way’s discs are. Here, we conduct an integral field spectroscopy study of the thin and thick discs in the edge-on Milky Way analogue IC 2531, using observational data from the WiFeS IFU spectrograph on the ANU 2.3m telescope. We provide spectral analysis of IC 2531’s kinematics and stellar populations in the thin and thick disc regions by conducting \textsc{ppxf} fitting with Vazdekis models. We found that IC 2531’s disc above the dust plane generally has chemical properties between the Milky Way’s chemical thin and thick disc. IC 2531’s thick disc is somewhat similar to the Milky Way’s in terms of stellar populations: on average, it is older, more metal-poor than its corresponding thin disc, but similarly alpha-enhanced in general. But we do find a clearly alpha-rich thick disc bin though one of our “thick disc” bins may be dominated by a warped thin disc. These results may help to constrain formation theories of thick discs.
\end{abstract}


%%%%%%%%%%%%%%%%%%%%%%%%%%%%%%%%%%%%%%%%%%%%%%%%%%

%%%%%%%%%%%%%%%%% BODY OF PAPER %%%%%%%%%%%%%%%%%%

\section{Introduction}\
\label{sec:1}

One of the unsolved problems in galaxy evolution is the formation of thick discs in spiral galaxies. We know that most large spiral galaxies consist of geometrical thin and thick discs, but we still do not have a clear understanding about how they form. The Milky Way's discs have been studied in relative detail and there is a range of theories of their origin. Thin and thick discs are different chemically: thin and thick disc stars near the sun have different ages, metallicity [M/H] and [$\alpha$/Fe] distributions. Thin disc stars are younger, metal-rich and alpha-poor; thick disc stars are older, metal-poor and alpha-rich, e.g.~\citet{2013A&A...560A.109H, 2014A&A...562A..71B}. The thin disc’s low-alpha value means that its enrichment was gradual, mainly on SNIa timescales (believed to be $\sim$ a few Gyr), and the star formation duration is long. The thick disc’s chemical evolution time is short compared to the timescale, for the SNIa. This indicates that thin and thick disc stars originate at different phases of the galactic assembly process. On a larger (several kpc) scale, high alpha stars are found mainly at high $z$ (actual heights above the galaxy plane) and in the inner disc, and there is a negative radial metallicity gradient in the thin disc~\citep{2015ApJ...808..132H}. The inner galaxy has a similar distribution, with high alpha stars dominating at high $z$ and high metallicity stars dominating in some regions near the plane~\citep{2021A&A...653A.143W}. Several formation theories of the thick disc have been promoted, such as internal vertical heating~\citep{1985ApJ...290...75V}, merging of gas-rich galaxies~\citep{2004ApJ...612..894B}, accreted satellite galaxies~\citep{2003ApJ...597...21A}, or turbulent clumpy gas (upside-down formation) e.g.~\citep{2012ApJ...758..106K, 2015ApJ...799..209W}.

While these general properties of galactic discs have been known for many years, similar studies on other galaxies are still few in number and there is the concern that the Milky Way may be atypical. We can directly measure individual star’s age and chemical abundance in the Milky Way, but such studies had been out of reach in other galaxies until recently. We note that for galaxies within a few Mpc, low resolution spectroscopy of individual stars has become possible with integral field units (IFUs) on the largest telescopes, e.g.~\citet{2018A&A...618A...3R}. For more distant galaxies, the usual method for measuring stellar age and chemical abundance is via slitlet or integral field spectroscopy of the integrated light.

The first extensive kinematics study of thin and thick discs in other galaxies and comparison with the Milky Way’s was by~\citet{2005ApJ...624..701Y}. They were also the first to measure stellar population properties of thin and thick discs in other galaxies~\citep{2008ApJ...683..707Y}. Those two pioneer studies were done with conventional slitlet spectrographs and Lick indices. After that, only a few studies were attempted until the maturing of Integral Field Spectroscopy (IFS) and stellar population analysis techniques, which made it possible to measure kinematics and stellar population in a 2D field of view, e.g.~\citet{2015A&A...584A..34C}. With recent developments in stellar libraries, synthetic spectra and spectrum fitting codes, our ability to model galaxy spectra has advanced rapidly.~\citet{2015MNRAS.449.1177V} introduced a spectral library based on MILES~\citep{2006MNRAS.371..703S} spectra (referred as Vazdekis models in this paper), which is widely used in this kind of research.

\textsc{ppxf} is a widely used package for fitting spectra and deriving stellar parameters. It was first presented by~\citet{2004PASP..116..138C}, and was gradually improved over time especially in~\citet{2017MNRAS.466..798C}. \textsc{ppxf} is able to fit a large number of stellar templates to obtain excellent results on the kinematical, age, and abundance properties of stellar populations. Studies of other galaxies' thin and thick discs since 2016 have mostly used the combination of \textsc{ppxf} and Vazdekis models to fit observed IFS spectra. These new techniques allow us to study galaxies in a 2D field of view with a higher spatial resolution and better stellar population resolution. As a result, the amount of research in this area has increased significantly in the last decade~\citep{2015A&A...584A..34C, 2016A&A...593L...6C, 2016A&A...591A.143G, 2018A&A...616A.121S, 2019A&A...623A..19P, 2019A&A...625A..95P, 2021ApJ...913L..11S, 2021MNRAS.508.2458M, 2023MNRAS.520.3066S}.  Other options are long slit spectroscopy, NBURSTS~\citep{Chilingarian_Prugniel_Sil’chenko_Koleva_2006, 2007MNRAS.376.1033C}, and PEGASE-HR models~\citep{2004A&A...425..881L}, see~\citep{2016MNRAS.460L..89K, 2019MNRAS.483.2413K, 2020MNRAS.493.5464K}.

These tools have significantly improved our understanding in this field and have proved several conjectures: most large edge-on spirals appear more or less similar to the Milky Way. Similar stellar population differences between thick and thin discs are seen~\citep{2008ApJ...683..707Y, 2015A&A...584A..34C, 2016A&A...593L...6C, 2016A&A...591A.143G, 2019MNRAS.483.2413K, 2019A&A...623A..19P, 2019A&A...625A..95P, 2020MNRAS.493.5464K, 2021ApJ...913L..11S, 2021MNRAS.508.2458M, 2023MNRAS.520.3066S}, and some galaxies exhibit kinematics and rotational lag of thick discs similar to the Milky Way~\citep{2005ApJ...624..701Y,2015A&A...584A..34C,2016A&A...593L...6C}. Nevertheless, the number of edge-on galaxies for which we have detailed spectral studies done with IFUs remains small, which provides the motivation to conduct more similar observations to increase our sample size. 

In these observations, we have seen different formation rates of thin and thick discs in other edge-on disc galaxies, such as the fast and early formation of thick discs and the secular formation of thin discs later (IC 335, NGC 1380A, NGC 1381)~\citep{2019MNRAS.483.2413K}. In some systems both discs appear to form in a short period (ESO 533-4, ESO 243-49)~\citep{2015A&A...584A..34C,2016A&A...593L...6C}. In others we see the complete opposite, with both discs forming gradually~\citep{2016A&A...591A.143G}. These diverse results all suggest that the galaxy disc’s formation process may not be unique.~\citet{2019A&A...625A..95P} shows that the environment and interaction with other galaxies play an important role in galaxy evolution: galaxies located in a less dense environment may have longer star formation epochs than galaxies near a cluster centre. However, a galaxy’s position in a cluster changes with time and this phenomenon may not be very well defined, see~\citet{2016MNRAS.460L..89K} for counter-examples.

IC 2531 is an almost perfectly edge-on (inclination angle $i=89.5$°) Hubble-type Sc spiral galaxy in the Constellation Antlia. It lies in a relatively clean environment with only two dwarf companions discovered by H\,\textsc{i} observations~\citep{2004MNRAS.352..768K}. Past observations have shown that IC 2531 is comparable to the Milky Way in maximum rotation speed~\citep{2010A&A...515A..62O}, bar geometry~\citep{2010ASPC..424..261A}, and a modest star formation rate~\citep{2000A&A...359..433R}. A simulation~\citep{1996A&A...309..715J} suggests it can be modelled with stellar populations similar to the Milky Way. These similarities all make it an ideal Milky Way analogue. It is for these reasons that we chose IC 2531 as our target. The distance to IC 2531 is still uncertain: It is mainly measured via the Tully-Fisher relation: recent distance estimates in NASA/IPAC Extragalactic Database range from about 27 to 37\,Mpc. We adopt 32\,Mpc in this paper, so 1\,arcsec equals about 155\,pc.

We note that definitions of thick discs in literature are not unique. It can be defined as a chemical or geometrical thick disc and these two definitions do not quite overlap. The chemical definition is commonly used in studies of the Milky Way, and is determined by the alpha-enhancement of thick disc stars relative to the thin disc's. The geometrical definition is mostly used in external galaxies where we cannot resolve individual stars, and is defined as a low-surface brightness relatively red envelope with similar colour and geometry to the Milky Way’s thick disc~\citep{2002AJ....124.1328D,2019A&A...623A..19P}. Just as for thin discs, the density distribution of thick discs is usually fit by a double exponential function of the form  $(r/h_r+\lvert z \rvert / h_z)$, where $(r,z)$ is a position in the disc, $r$ and $R$ are its cylindrical and projected radii, $h_r$ and $h _z$ are the related disc’s scale length and scale height.

The structure of thick discs is usually measured with surface photometry, with limiting g-band surface brightness of 27-28 mag arcsec$^{-2}$. A few studies reach somewhat deeper:

(1) imaging of galaxies within about 10 Mpc with HST or with large ground-based telescopes and very long exposures can resolve the brightest giant stars in the outer regions of edge-on galaxies. Using star counts rather than surface brightness reduces the background noise and enables the detection of fainter structures with g-band surface brightness below 30 mag arcsec$^{-2}$. 

(2) instruments like DRAGONFLY (e.g. \citet{2022ApJ...932...44G}), which use deep imaging of diffuse light, are designed to minimise the contributions from scattered light and uncertainties from the sky background, enabling detection of fainter galactic structures with a g-band surface brightness of about 29 to 30 mag arcsec$^{-2}$ in long exposures.

\begin{figure}
    \centering
    \includegraphics[width=1.0\linewidth]{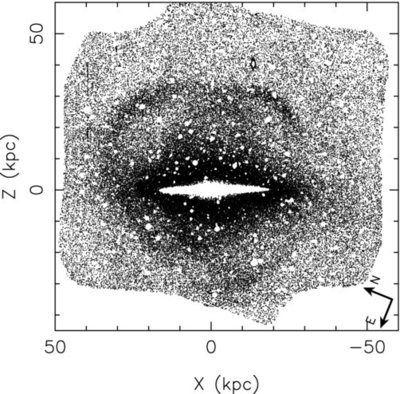}
        \caption{The diamond-shaped outer isophotes from Subaru star counts outline the Milky Way analog NGC 891's extended diamond-shaped structure (figure 1 of~\citet{2010ApJ...714L..12M}, we thank Prof. Rodrigo Ibata for allowing us to use this image in our paper).}
    \label{fig:ngc891}
\end{figure}

NGC 891, a nearby (9.1 Mpc) edge-on analog of the Milky Way, is an example. Its outer structure has been studied in several papers. \citet{2010ApJ...714L..12M} deep star count study of NGC 891 (see Figure~\ref{fig:ngc891}) with the Subaru Supreme-Cam shows an extended diamond-shaped structure enveloping the (white) thin disc. The diamond shape indicates an underlying disc structure with a surface brightness distribution that is a decreasing function of $(r/h_r+\lvert z \rvert / h_z)$. Their image provides a very deep visualization of such an extended disc structure.  The diamond-shaped structure extends to about $R = 30$ kpc, $\lvert z \rvert$ = 15 kpc, well beyond the double-exponential thick disc ($h_z$ = 1.44 kpc, $h_r$ = 4.8 kpc) detected in HST/ACS star counts by \citet{2009MNRAS.395..126I}: this thick disc extends to only about $\lvert z \rvert$ = 5 kpc. \citet{2010ApJ...714L..12M} point out that (1) the large diamond-shaped envelope is unlike any classical thick disc or halo, and may be the debris of accreted satellites; (2) the galaxy NGC 2683 has a similar vertically extended structure.

In their survey of edge-on galaxies, \citet{2020MNRAS.494.1751M} recognise the discy diamond shape as one of three different morphologies of the outer light distribution in edge-on disc galaxies: oval and boxy are the other two. Eight of their 35 galaxies have the diamond-shaped morphology (their photometry for NGC 891 extends out to a radius of about 21 kpc: cf. Figure~\ref{fig:ngc891}).

Although our present IFU study of the edge-on galaxy IC 2531 cannot reach this kind of very faint outer structure, the point of these comments is to caution that there may be more to these discs than the thin and thick discs.

Here we are interested primarily in chemically-defined thin and thick discs, but we can only work on geometrically-defined thin and thick discs until the results come out: we bin our data geometrically, and study their stellar population, to see whether or not IC 2531’s geometrical thin and thick discs have chemical properties that are similar to the Milky Way’s chemical thin and thick discs. In this paper, thin and thick discs by default mean geometrical thin and thick discs, unless specifically specified to be “chemical”.

The goal of this project is to use integral field spectroscopy and synthesis spectra fitting techniques to probe kinematics and stellar population features in the Milky Way analogue IC 2531. It is an ideal galaxy to address the question of whether the Milky Way's characteristics and history are typical for other large spiral galaxies.

The layout of this paper is as follows. In Section~\ref{sec:2}, we describe our observations and data processing. The results are provided in Section~\ref{sec:3}. Section~\ref{sec:4} provides a summary.

\section{Observation and data analysis}
\label{sec:2}

In this section we describe our observations and data analysis, as well as the instrument and software involved. We observed two positions in IC 2531 with the IFU WiFeS on the Australian National University (ANU) 2.3m telescope in nod-and-shuffle mode. 
After reducing the data, we process and spatially bin the data to provide adequate SNR (Signal to Noise Ratio) before conducting spectral fitting.

\subsection{Observations}
\label{sec:maths} % used for referring to this section from elsewhere

Here we describe the observations and the IFU instrument WiFeS.

Observations were carried out on the ANU 2.3m telescope with WiFeS, an IFU spectrograph developed by~\citet{2007Ap&SS.310..255D}. It is fed by an image slicer which functions as a multislit with near-zero dead space. The 25 multislits are each 1" wide on the sky and are contiguous. The multislit length is 38" so the device covers a 25$\times$38" region of the sky with 1"$\times$1" pixels.

\begin{figure}
    \centering
    \includegraphics[width=1.0\linewidth]{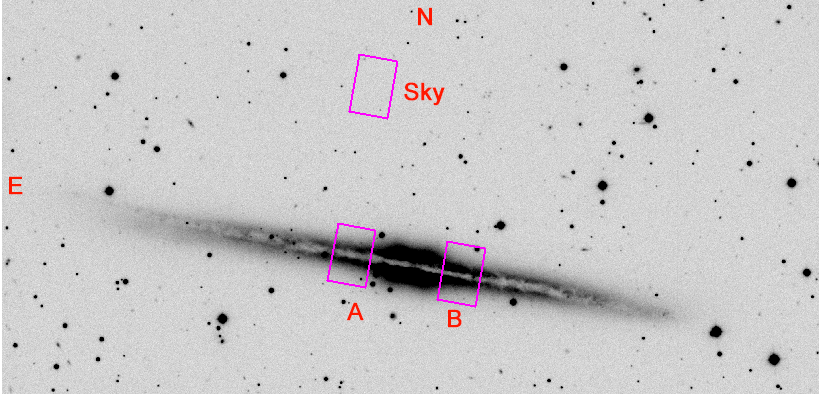}
        \caption{Pos A and pos B in IC 2531, plotted on top of a V band image from the Carnegie-Irvine Galaxy Survey (CGS), the size of the IFU is 25$\times$38". Our sky location for the nod-and-shuffle observation is also shown. We thank Prof. Luis C. Ho and the Carnegie-Irvine Galaxy Survey team for allowing us to use their images.}
    \label{fig:posab}
\end{figure}

To analyse the kinematics and stellar populations of the thin and thick disc components of IC 2531, we collect deep spectra on two different positions, referred to as pos A and pos B (see Figure~\ref{fig:posab}), with centres at J2000 (09:59:58.4. --29:36:52), and (09:59:53.0, --29:37:09) respectively. Pos A and pos B both cover the galaxy mid-plane (thin disc) and the area above the mid-plane, where the thick disc is likely to prevail. The projected radii of centres of pos A and pos B are about 40" and 32" respectively; at the adopted distance of 32 Mpc, pos A covers a projected radius of $4.3-8.1$ kpc, pos B covers a projected radius of $3.1-6.8$ kpc, with each position covering 2.9 kpc above and below the galactic central plane. In comparison, ~\citet{2016A&A...592A..71M} gives the scale lengths of the thin and thick discs of IC 2531 are 6.96 and 21.63 kpc respectively (these scale lengths are very large compared to the 2.6 and 2.0 kpc scale lengths of the Milky Way's thin and thick disc \citep{2016ARA&A..54..529B}), and the scale heights are 0.53 and 1.37 kpc respectively (adjusted to our adopted distance of 32 Mpc). So radially, pos A and B center at 0.89 and 0.71 thin disc scale length; vertically, they cover over two times the thick disc’s scale height and five times the thin disc’s scale height.

WiFeS is a double-beam spectrograph with red and blue arms, and a choice of grating and dichroic beam splitters. For these observations, we use the B7000 grating with a bandpass of $4170 - 5550$\,\AA\, and a 2-pixel resolution of $R = 7000$. The observing mode we use was nod-and-shuffle, which gives sky subtraction with noise close to the photon and readout noise limit. The telescope is nodded between galaxy and sky fields (see Figure~\ref{fig:posab}), and both galaxy and sky spectra are shuffled and accumulated on the CCD for several cycles of nod-and-shuffle. Each individual observation used 6 cycles of 150s exposure alternately on galaxy and sky, reading out after 1800s of integration. The total observing time was 315 minutes on each galaxy position (A and B) with the same for the sky.

Ne-Ar arc lamp exposures were used to calibrate the wavelength scale, and quartz lamp flats were taken regularly to calibrate for pixel-to-pixel variations. Twilight flats and bias frames were also taken. HD 52265 was observed as the radial velocity standard.

The product of observations in each of the two galaxy positions is a data cube with $25\times38$ 1"$\times$1" pixels on-sky and $3971\times0.35$\,\AA\, pixels in wavelength.

\subsection{Data reduction}
We reduced WiFeS data using a modified version of the public pipeline \textsc{pywifes}~\citep{2014Ap&SS.349..617C}. In brief, \textsc{pywifes} repairs bad pixels, reduces the bias, and cuts each spectrum into 25 individual slitlet spectra before performing cosmic ray reduction and sky subtraction. Flat field and wavelength calibrations are applied after multiple spectra are combined. Finally, the 3D data cube ($25\times38\times3971$ pixels) is constructed.

Some of the spectra in pos A were taken in enhanced solar activity. Despite the benefit of nod and shuffle sky subtraction, the geocoronal H$\upbeta$ sky line at zero redshift in these spectra was not fully subtracted. We were unable to determine why sky subtraction did not work perfectly for these spectra. We looked for evidence of solar (absorption line) contamination of galaxy spectra at zero redshift but none was apparent.

\subsection{Image reconstruction}

Here we describe our data processing steps, which include combining multiple observations (scaling, median-combining), Gaussian smoothing, and 2D image reconstruction.

For each position, we calculated mean counts $A_i$ of each reduced 3D data cube, and calculated the overall average $A_m$ of all the mean counts $A_i$. Then each spectrum is scaled by multiplying a factor $A_m$/$A_i$ to the same brightness (mean counts) level. After that, they were median-combined. Then spectra were Gaussian smoothed along the wavelength axis to increase its FWHM from $0.7\,$\AA\ to $2.51\,$\AA, to match the resolution of the template MILES~\citep{2006MNRAS.371..703S} spectra in the subsequent spectral fitting process. 

\begin{figure*}
    \centering
    \includegraphics[width=1.0\linewidth]{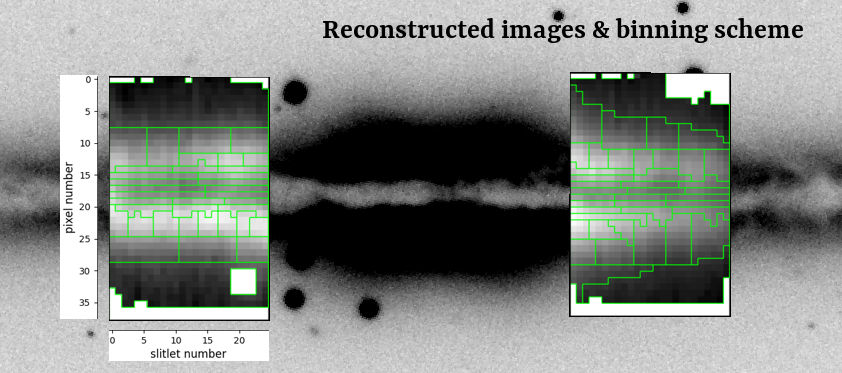}
        \caption{Reconstructed images for pos A (left) and pos B (right) with our binning scheme (green, to be discussed in section~\ref{sec:2.5}) overlaid on the CGS V band image of IC 2531. Our reconstructed images are in grayscale, and the CGS image is in reverse grayscale. The bright parts in our reconstructed images and the black parts in the CGS image represent the geometrical thin disc, and their outlines are continuous. The white regions represent bad pixels and areas contaminated by foreground stars.}
  \label{fig:cjtx}
\end{figure*}

As a check on the precise observational positions, we reconstructed galaxy images using the reduced data cubes. We took a spectral region with a wavelength range of $5450.2$\,\AA\, to $5500.2$\,\AA\,, a region near the red end of the spectrum with higher counts and has no strong emission or absorption features. Figure~\ref{fig:cjtx} shows grayscale reconstructed images of our two IFU fields on top of the CGS V band image of IC 2531, where each pixel corresponds to a 1"$\times$1" region of the sky. The reconstructed images match the direct optical image of IC 2531 quite well, including details such as changes in the shape of the dust lane. 

However, after reduction, a few bad pixels still appear in rows near the top and bottom in some of the slitlets. These bad pixels and two foreground stars (cf. Figure~\ref{fig:posab}) can be seen as the white regions that span 2-4 pixel rows at the high and low $y$-values in reconstructed $x$-$y$ images (Figure~\ref{fig:cjtx}) and have a minor effect on our binning strategy in section~\ref{sec:2.5}.

In Figure~\ref{fig:cjtx} and throughout this paper, $x=1:25$ represents the slitlet number, which is parallel to the major axis of IC 2531. $y=1:38$ represents the pixel number along the slitlet, which is parallel to the minor axis of IC 2531. Actual heights above the galaxy plane are referred to by the conventional $z$.

\subsection{Scale-height discussion}
\label{sec:2.4}

\begin{figure*}
    \centering
    \begin{minipage}{0.33\linewidth}
        \centering
        \includegraphics[width=1\linewidth]{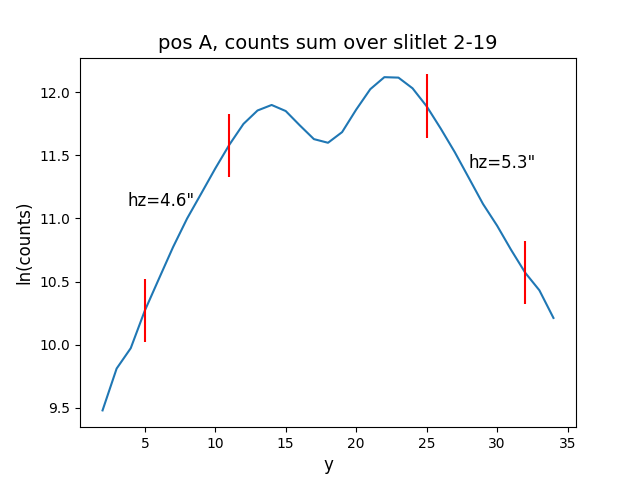}
    \end{minipage}
    \begin{minipage}{0.33\linewidth}
        \centering
        \includegraphics[width=1\linewidth]{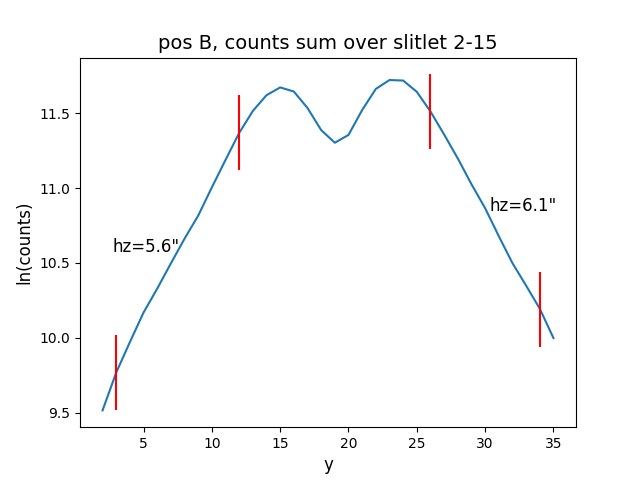} 
    \end{minipage}
    \begin{minipage}{0.33\linewidth}
        \centering
        \includegraphics[width=1\linewidth]{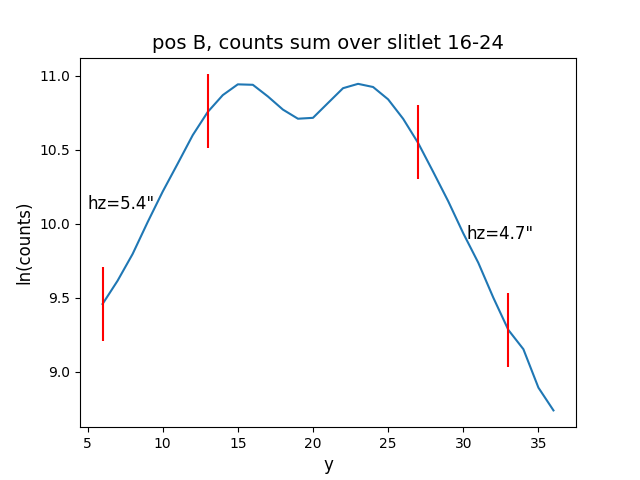}
    \end{minipage}
    \caption{Left panel: Plot of the vertical light (wavelength 5450 to 5500\,\AA) distribution in pos A of IC 2531, averaged over the radial zone $R=29$" to 46" (4.5 to 7.1\,kpc). The short vertical red lines indicate the bright limit of the exponential outer structure of the vertical light distribution used for estimating the scale height. Middle panel: similar to the left panel, but for pos B-1, averaged over the radial zone $R=21$" to 34" (3.3 to 5.3 kpc). Right panel: similar to the left panel, but for pos B-2, averaged over the radial zone $R=$35" to 43" (5.4 to 6.7 kpc).}
  \label{fig:m shape}
\end{figure*}

We plot the surface brightness profiles parallel to the minor axis of IC 2531 in Figure~\ref{fig:m shape}. One can clearly see that the data gives M-shaped light distributions, which agrees well with profiles in figure 6 (b)\&(c) of~\citet{1989ApJ...337..163W}. The depression in the middle of the “M” is caused by the dust lane which obscures the starlight behind it. As the height above the central plane increases, the dust gets thinner and becomes more transparent. Beyond about 3.5" above and below from the central plane, the dust extinction is minimal and dust is almost transparent (cf. K band data in figure 10 (c)\&(d) of~\citet{1989ApJ...337..163W}).

In pos A, the southern side of the galaxy (corresponding to the bottom of all images presented in this paper) is significantly brighter than the northern side, which may be caused by an asymmetrical distribution of dust. The brightness of pos B's northern and southern sides are roughly the same.

The vertical light distribution of a single-component galaxy disc at a certain radius is often represented as $L_z=L_0 \exp(- \lvert z \rvert / h_z)$~\citep{1989ApJ...337..163W}. The combined light profile of a thin and thick disc can therefore be expressed as $L_z=L_{0t} \exp(- \lvert z \rvert / h_{zt})+ L_{0T} \exp(- \lvert z \rvert / h_{zT})$, where $T$ represents the thick disc component and $t$ represents the thin disc component. Therefore, in a plot of $\log(L)$ vs. $z$, the profile slope reflects the scale height $h_{zt}$ for small $\lvert z \rvert $ and tends to reflect the scale height $h_{zT}$ at larger $\lvert z \rvert $. The negative slope of the thin disc component's vertical light profile would be steeper as $ h_zt < h_zT $, resulting in an inflection point in the overall light profile of a two-component disc galaxy, representing the transition between regions dominated by thin and thick discs respectively, as shown in \citet{2015A&A...584A..34C}. 

However, the inflection point is absent in our data, and we see only a single straight-line component. We note that an analysis of IRAC 3.6 $\mu$m data by~\citet{2016A&A...592A..71M} shows inflection points at about 15" above the galactic plane, which is near the edge of our fields. In these regions, M curves of individual slitlet show fluctuations that may be affected by bad pixels at the top and bottom of some slitlets. Therefore, the lack of inflection points in our observed vertical light profiles may be due to errors in counts in those areas.

In pos A, for the light profile summed over slitlets 2 to 19, all of which have similar shapes, the upper N (northern) and lower S (southern) parts of the profile are close to exponentials with scale heights $h_z$ of 4.6" and 5.3" for the N and S sides respectively, as shown in in the left panel of Figure~\ref{fig:m shape}. The middle and right panels of Figure~\ref{fig:m shape} show the vertical light distribution profiles in pos B. In pos B, for slitlets 2 to 15 (referred as pos B-1), the apparent scale heights are 5.6" and 6.1" for the N and S sides respectively. For slitlets 16 to 24 (referred as pos B-2), the apparent scale heights are smaller (5.4" and 4.7"); this region lies at a larger radius, see Appendix \ref{appendix:B}, where we provide a more detailed analysis of scale heights in different regions of IC 2531 and discussed whether or not our fields are significantly contaminated by the bulge (no).

Compared with previous papers modeling IC 2531 (adjusted to our adopted distance), our apparent scale heights above are similar to scale heights of \citet{2016A&A...592A..71M}’s observed data (about 5") and also to those models with only one disc component \citep{1999A&A...344..868X, 2002MNRAS.334..646K}. The modeled galaxy simulated by \citet{2016A&A...592A..71M} contains a thin disc with a scale height of about 3.4" and a thick disc with a scale height of about 8.8", and shows inflection points at heights of about 14" (south) and 15" (north). These comparisons suggest the majority of vertical disc profiles in our fields are combinations of thin and thick discs. Regions just above and below the dust lane are thin disc dominated; the thin disc contribution decreases and the thick disc contribution increases with height, and they eventually become thick disc dominated at the top and bottom of our fields.

In summary, we see only a single exponential component with an apparent scale height between 4.6-6.1", corresponding to about 740-980\,pc , which is similar to the Milky Way's thick disc scale height of about 900\,pc (the Milky Way's thin discs' scale height is about 300\,pc) \citep{2016ARA&A..54..529B}. But comparison with modeled scale heights from previous researches suggests the majority of vertical disc profiles in our IC 2531 fields are a combination of the thin and thick discs.

\subsection{Spatial binning}
\label{sec:2.5}

\begin{figure*}
    \centering
    \includegraphics[width=1.0\linewidth]{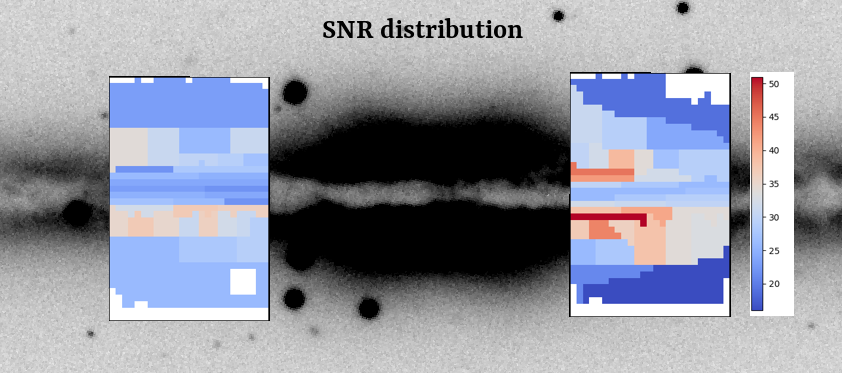}
        \caption{Plot of our bins' SNR distribution. The SNR of most of our bins are between 20 to 40. The narrower dust lane and thin disc bins still have adequate SNR despite being small, as they are in bright parts of IC 2531.}
  \label{fig:SNR}
\end{figure*}

Instead of using the popular Voronoi tessellation~\citep{2003MNRAS.342..345C}, we chose to bin our data to follow the spatial distribution of physical properties. We tried to use square bins to achieve as high a resolution as possible in both spatial directions on the sky, but in most cases we had to expand them to rectangular or irregular shapes to achieve adequate SNR for our analysis. The dust lane was primarily binned according to different heights in order to account for the fact that the mean depth along the line of sight depends on the dust's optical depth, which in turn varies with height. The SNR of a bin is calculated as the ratio between the average of \textsc{ppxf} stellar fit and the standard deviation of residuals, in the wavelength range 4950-5200\,\AA\, of its spectrum. We adjusted bin sizes with the aim of achieving a $\rm SNR>20$ for individual bins to get more precise results from \textsc{ppxf}, so that bins in the brightest part of the galaxy (low $z$, near the galactic nucleus) are small, while bins in fainter regions are much larger. Due to the thick disc being faint in nature, we only obtain four relatively low-SNR bins in the geometrical thick disc region. As shown in Figure~\ref{fig:SNR} and listed in Tables~\ref{tab:MCA} and~\ref{tab:MCB}, spectral quality obtained from this binning strategy is satisfactory and quite consistent across both fields; the calculated SNR of the bins’ spectra are generally between 20 to 40, with a few reaching 40 to 50. The spectra of the faint thick disc bins are noisier (SNR $\approx$ 20), but still adequate for \textsc{ppxf}. 

We show our binning scheme overlaid on reconstructed galaxy images in Figure~\ref{fig:cjtx}. Bin maps presented in Figures~\ref{fig:binac} and~\ref{fig:binbc} show the detailed bin distribution in each position (as well as fitting results as discussed in Section~\ref{sec:3}). The irregular bin shapes at the edge of the dust lane are due to different thickness of the dust at different radii. The irregular bin edges at the top and bottom of the images are due to the reduction issues and foreground stars; the white regions are designated as “bad” and “star” in bin maps and are excluded from our analysis. Please note that the colours in these two figures do not have any physical meaning and simply represent different bins.

We have 78 bins in total, comprising 37 in pos A and 41 in pos B. Our naming convention is as follows: “AT” and “BT” refer to bins we consider to be in the geometrical thick disc region that are faint and above the bright edge-on galaxy plane. “A…t…” and “B…t…” are bins we consider to be in the geometrical thin disc region that makes up the bright edge-on galaxy plane. “A…d…” and “B…d…” represents bins that are affected by the dust lane (i.e., regions between the peaks of the M-shaped light distribution shown in Figure~\ref{fig:m shape}). These designations are followed by the coordinate range of a bin in $x$ and $y$: e.g. A1-6t9-15 corresponds to a thin disc bin in pos A with coordinates ranging from 1 to 6 in $x$ and 9 to 15 in $y$. ($x=1:25$ represents the slitlet number, and $y=1:38$ is the pixel number along slitlet perpendicular to the disc plane, where zero is located at the top of the image). We note again that the large and small scale structure of our binning system is designed to follow the morphology of IC 2531 within our two IFU fields. We list the arithmetic mean projected radius and mean height of each bin in Tables~\ref{tab:MCA} and~\ref{tab:MCB}, zero of the height is taken as the arithmetic mean of dust lane pixels in each position, which are $y=18$ in pos A and $y=20$ in pos B.

\subsection{Spectral fitting}

We fitted the observed spectra with \textsc{ppxf}~\citep{2004PASP..116..138C, 2017MNRAS.466..798C}. Given an input galaxy spectrum, \textsc{ppxf} finds the best-fitting linear combination of a set of input stellar templates spanning various stellar population parameters, whilst simultaneously fitting for stellar kinematics. \textsc{ppxf} optionally also fits emission lines to provide fluxes and gas kinematics. 

We used the Vazdekis stellar  templates~\citep{2015MNRAS.449.1177V} which are based on the empirical MILES library~\citep{2006MNRAS.371..703S} and are generated using models based on BaSTI stellar isochrones. We used the templates corresponding to a Kroupa Universal initial mass function with a slope of 1.3. 
We included 672 templates with 28 logarithmically-spaced ages from 0.5 to 14 Gyr, 12 values of metallicity [M/H] from $-2.27$ to 0.4, and the 2 available values of [$\alpha$/Fe] of 0 and 0.4. These models cover the wavelength range 3540-7410\,\AA, and are provided at a spectral resolution of 2.51\,\AA\, (FWHM), and sampled at 0.9\,\AA/pixel. 
In addition to the SSP template spectra described above, we also included four ionised gas emission lines within our wavelength range which covers ${\rm H}\upgamma$, ${\rm H}\upbeta$, and the [O\,\textsc{iii}] doublet at 4959\,\AA\, and 5000\,\AA. By fitting stellar and gas templates simultaneously, we do not need to mask wavelength ranges that may contain important emission and absorption lines like ${\rm H}\upbeta$, the most useful absorption-line age indicator in our wavelength range, which is also present as an emission line from the ionised gas in H\,\textsc{ii} regions.

The kinematic parameters fitted by \textsc{ppxf} include the average line-of-sight velocity, velocity dispersion, $h3$, and $h4$, where these last two parameters are Gauss-Hermite moments and represent the skewness and kurtosis of the line-of-sight velocity distribution. All emission lines are fitted simultaneously as simple Gaussian profiles, with the line-of-sight velocity and velocity dispersion being fixed between all emission lines but amplitudes allowed to vary.
Stellar population parameters included in the fit are $\log_{10}$(\rm{age}\,[Gyr]) - hereafter abbreviated as lg(age), [M/H], and [$\alpha$/Fe]. \textsc{ppxf} assigns mass weights to each template to obtain the best fit to the observed spectrum in each bin, from which we can calculate the mass-weighted mean of each population parameter.

\begin{figure}
    \centering
    \begin{minipage}{0.49\linewidth}
        \centering
        \includegraphics[width=1\linewidth]{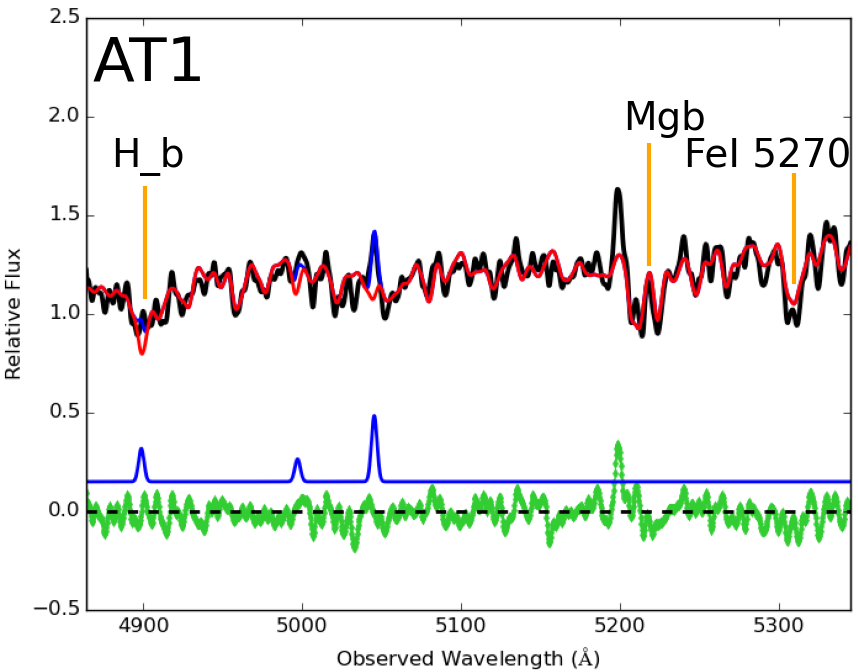}
    \end{minipage}
    \begin{minipage}{0.49\linewidth}
        \centering
        \includegraphics[width=1\linewidth]{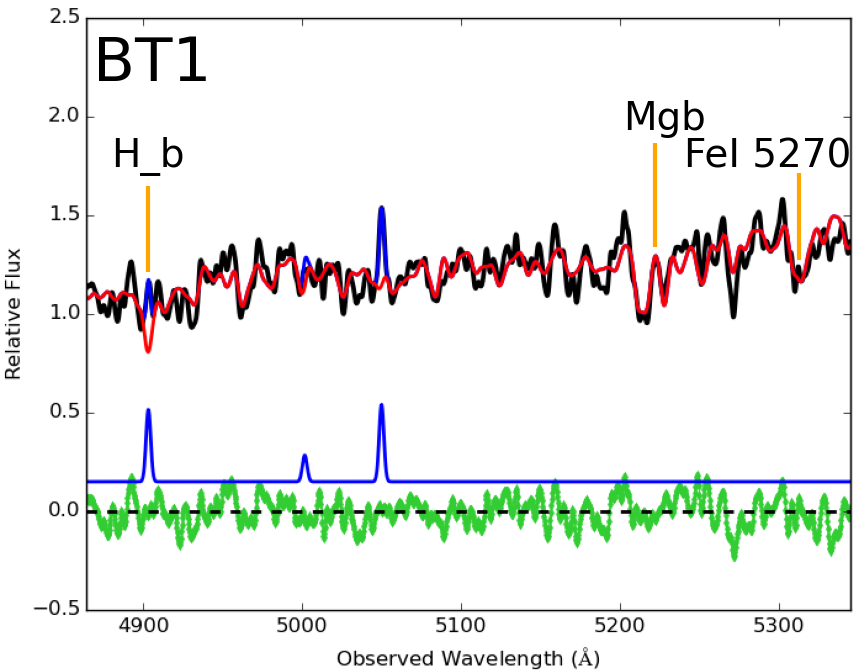}
    \end{minipage}
    \begin{minipage}{0.49\linewidth}
        \centering
        \includegraphics[width=1\linewidth]{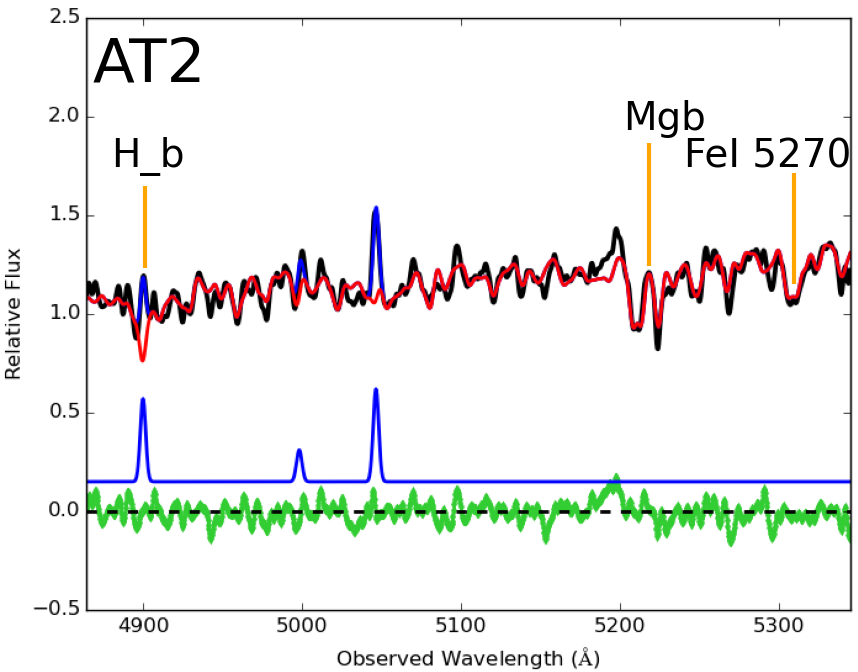}
    \end{minipage}
    \begin{minipage}{0.49\linewidth}
        \centering
        \includegraphics[width=1\linewidth]{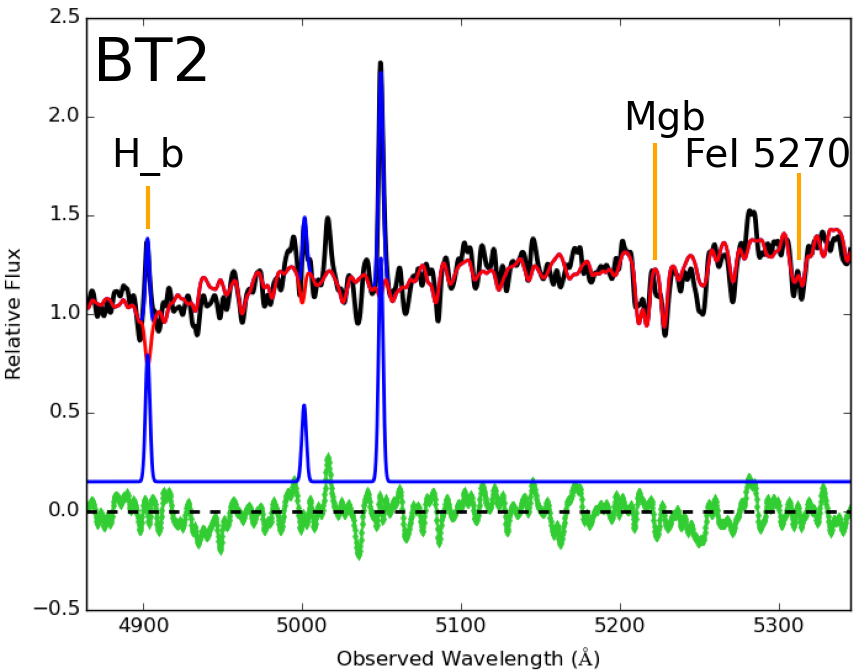}
    \end{minipage}
    \caption{The 4864-5345\,\AA\, portion of the \textsc{ppxf} fitting results for four thick disc bins. Top left panel: normalised integrated spectrum for AT1 in black overlaid by the \textsc{ppxf} best fit in red. Ionised gas emission lines (${\rm H}\upbeta$, [O\,\textsc{iii}]) are shown in blue. Residuals are shown in green. Bottom left panel: same as above but for AT2. Top right panel: BT1's. Bottom right panel: BT2's.}
  \label{fig:ftr}
\end{figure}

As an example, Figure~\ref{fig:ftr} shows the 4864-5345\,\AA\, portion of the \textsc{ppxf} fitting results for four thick disc bins, a range most relevant to our discussion and to be revisited in section~\ref{sec:3.3}. It can be seen that AT1's spectrum shows a much wider ${\rm H}\upbeta$ absorption line and stronger Fe\,\textsc{i} 5270 lines compared to the other three thick disc bins, indicating it is dominated by a younger population and has a higher [M/H], as estimated by \textsc{ppxf}.

\section{Results}
\label{sec:3}

In this section we present our kinematics and stellar population results from our \textsc{ppxf} fitting. Due to line-of-sight projection effects, the estimated stellar kinematics provide limited useful information, but can provide some insights about how far we are looking into the galaxy along each line-of-sight. Meanwhile, our stellar population fitting results reveal that IC 2531’s disc above the dust plane generally has chemical properties between the Milky Way’s chemical thin and thick disc. Of our four geometrical thick disc bins, we found two alpha-enhanced thick disc bins (AT2, BT1), one alpha-rich chemical thick disc bin (BT2) and one alpha-poor (AT1) which we tend to believe due to a warped thin disc.

Due to dust attenuation near the galaxy mid-plane, our observations only probe light from stars closest to us along the line-of-sight. As the line of sight moves away from the mid-plane to higher $z$, the dust's optical depth decreases, meaning our observations probe further into the disc, providing us with a view of the radial properties of the galaxy.

The discussion in Sections~\ref{sec:3.1} and~\ref{sec:3.2} will not include bin AT1 due to its unexpected properties; this bin is discussed further in Section~\ref{sec:3.3}.

The \textsc{ppxf} fitting results of each bin are summarised in Tables~\ref{tab:MCA} and~\ref{tab:MCB}, as well as in Figures~\ref{fig:binac} and~\ref{fig:binbc}: values in each bin after its name correspond to the mean stellar velocity, stellar velocity dispersion, ionised gas velocity, lg(age), [M/H], and [$\alpha$/Fe].

\subsection{Stellar Kinematics}
\label{sec:3.1}

Stellar velocities for an edge-on galaxy are not straightforward to interpret due to line-of-sight projection and dust obscuration: they cannot be interpreted as actual velocities at a given $(r,z)$ position in the galaxy, as is possible in less inclined galaxies, but rather give the dust modulated (when at low $z$) luminosity-weighted mean velocity for a projected position $(R,z)$.

It is well known that the rotation curve of a spiral galaxy is solid-body-like in its central region before it finally becomes flat at larger radii. The contours of constant radial velocities in the galaxy plane are known as the spider diagram: for a certain projected radius, the maximum velocity is at the line of nodes (i.e., the line of intersection of the galaxy plane and the sky plane), and decreases as the line-of-sight moves away from the line of nodes, e.g. see figure 8 of~\citet{2011A&A...528A..88G}. This information is useful in interpreting kinematic results.

In the following subsections we display maps of mean stellar velocity ($V$), stellar velocity dispersion ($\sigma$) and ionised gas velocity derived from our spectra via \textsc{ppxf}. Our velocities are heliocentric. Our study is aimed primarily at estimating stellar population parameters in inner thick disc regions at a few kpc above the galactic plane, as we do not have extended radial coverage. We recognise the importance of kinematic parameters $V$, $\sigma$, $h3$ and $h4$, e.g.~\citet{2005ApJ...626..159B, 2015MNRAS.450.2514I}, but for our purposes, kinematics are primarily a useful diagnostic to aid in interpreting the effects of line-of-sight projection and dust absorption. Whilst the parameters $V$ and $\sigma$ are useful for this purpose, we do not present our $h3$ and $h4$ results as interpreting these parameters at the SNR of our spectra is complex, and they are not critical for our purposes. We compare our optical velocities (stellar +
ionised gas) to H\,\textsc{i} velocities (for which the galaxy is often assumed to be optically thin) to understand how much of the lines-of-sight in different regions of the galaxy are effectively probed by our observations.

\subsubsection{Stellar velocity}

\begin{figure*}
    \centering
    \includegraphics[width=0.9\linewidth]{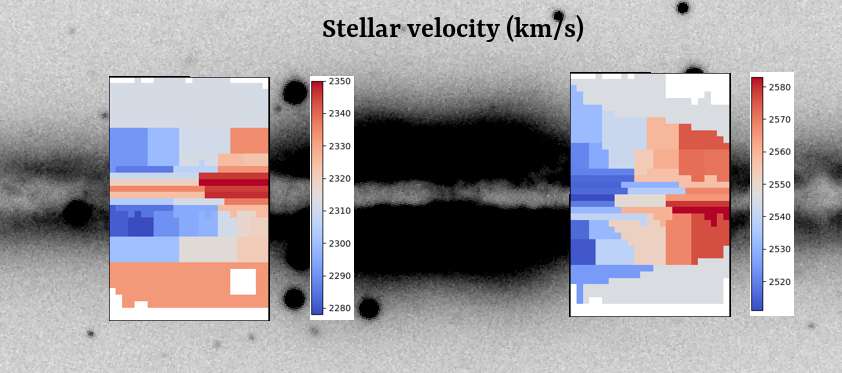}
        \caption{Maps of the line-of-sight stellar velocity fields for pos A and B overlaid on the CGS image of IC 2531. Note that we plot stellar velocities for pos A and B with different velocity scales to show more detailed velocity differences within each position; as an edge-on galaxy's velocity field is dominated by rotation, velocity differences within each position would be difficult to discern on the same velocity scale.}
    \label{fig:ke}
\end{figure*}

Figure~\ref{fig:ke} displays the stellar velocity map, which is typical of an edge-on galaxy's velocity distribution: the stellar velocities in the thin disc and dust lane show a positive gradient with projected radius, and stellar velocities on the north and south sides of the galaxy (above and below the plane in our images) are basically symmetrical, except that AT2's stellar velocity is $20 \, \rm km \, s^{-1}$ higher than AT1.

The mean radial stellar velocities of IC 2531 in pos A and pos B are approximately 2314\,$\rm km \, s^{-1}$ and 2544\,$\rm km \, s^{-1}$ respectively. Thick disc bins BT1 and BT2 both show a small rotation lag about 7\,$\rm km \, s^{-1}$ compared to the thin disc bins in the middle ($x \approx 13$) of pos B. There is a larger lag of about 30\,$\rm km \, s^{-1}$ in AT2 (noting that in pos A, low radial velocity corresponds to faster rotation), but AT1 shows no lag.

\begin{figure}
    \centering
    \includegraphics[width=1.0\linewidth]{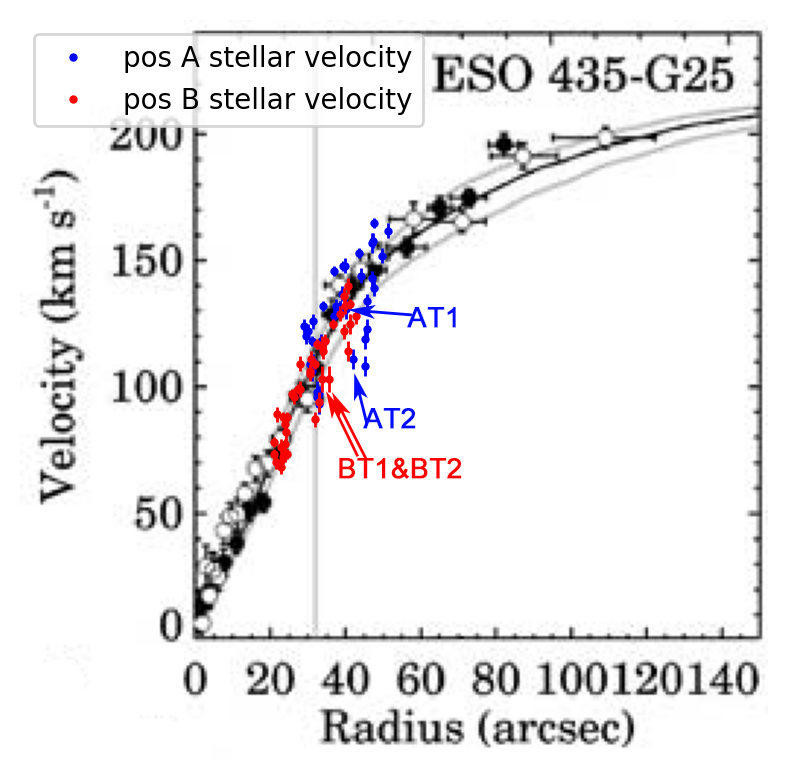}
        \caption{Plot of our stellar velocities over the observed stellar rotation curve from \citet{2005MNRAS.358..481K}, errors in this plot are \textsc{ppxf} errors. We thank Prof. Piet van der Kruit for allowing us to use this image in our paper.}
    \label{fig:orc star}
\end{figure}

We did a simple linear fit of all our $(p,v)$ data points (position, velocity), which gives a systemic velocity of 2443\,km\,s$^{-1}$, then we calculated the rotational stellar velocity of each bin. We then plot our rotational stellar velocity data, folded in radius, on top of the observed rotation curve of IC 2531 reported by \citet{2005MNRAS.358..481K} in Figure~\ref{fig:orc star}. Most of our velocities agree well with the observed rotation curve. Those below are mostly dust region bins and thick disc bins, but AT1 is right on the curve; Those above the curve are mostly thin disc bins at low $z$.

We assume that, in the mean, dust attenuation should decrease with height, and our line-of-sight should reach deeper into the galaxy. In the outer parts of the dust lane in pos A ($x$ from 1 to 15), the radial stellar velocity decreases as height increases, the dust becomes thinner, and we look deeper into the galaxy. This indicates that IC 2531's edge-on rotation curve reaches its flat part at approximately this radius; this phenomenon doesn't appear in the inner parts of the dust lane in pos A or in the pos B dust lane, which suggests that they lie on the linear part of the edge-on rotation curve (cf. Figure~\ref{fig:orc star} which shows pos B’s plots are on the linear part of the observed rotation curve, while pos A’s plots reach where the curve starts to turn flat).

We note that, assuming observed materials are in circular motion on average, it would be possible to derive line-of-sight depth for each bin, as it was shown in \citet{2018ApJ...853..114E}. However, the best-fitting model from \citet{2005MNRAS.358..481K} suggests that IC 2531's intrinsic rotation curve is nearly linear (where this method doesn’t work as the spider diagram is straight in this part) with radius out to about $r=50$". As our observation cover a projected radius from 20 to 52", this method doesn’t work for most of our bins because the contours of constant line of sight velocity in the spider diagram are straight (along the line of sight in an edge-on system). However, the outer half of pos A dust region may be in the flat part of IC 2531's actual rotation curve due to their larger actual radii, as discussed in the last paragraph. We calculate those dust region bins’ actual radii, which are listed in Table~\ref{tab:dpj}.

\begin{table*}
  \centering
  \caption{Calculated deprojected radii of dust region bins in pos A outer half, which may be in the flat part of IC 2531's actual rotation curve.}
    \begin{tabular}{cccc}
    \hline
          & projected $\lvert R \lvert$"  &  rotational star $v$ (km\,s$^{-1}$)    & deprojected $\lvert R \lvert$" \\
    \hline
    A2-10d15 & 47.0  & 157   & 65.9 \\
    A1-14d16 & 45.5  & 134   & 74.7 \\
    A1-14d17 & 45.5  & 123   & 81.4 \\
    A1-15d18 & 45.0  & 108   & 91.7 \\
    A1-15d19 & 45.0  & 119   & 83.2 \\
    A1-10d20 & 47.5  & 139   & 75.2 \\
    A2-10d21-22 & 47.1  & 158   & 65.6 \\
    \hline
    \end{tabular}%
  \label{tab:dpj}%
\end{table*}%

\subsubsection{Stellar velocity dispersion}

\begin{figure*}
    \centering
    \includegraphics[width=0.9\linewidth]{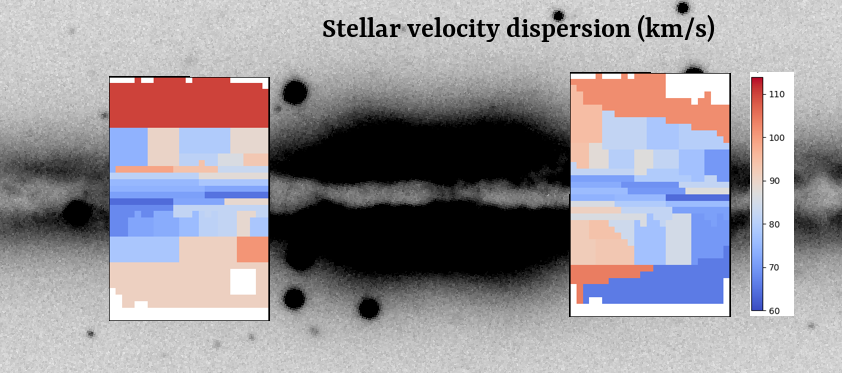}
        \caption{Similar to Figure~\ref{fig:ke}, but for stellar velocity dispersion fields.}
    \label{fig:dv}
\end{figure*}

Figure~\ref{fig:dv} displays the map of velocity dispersion. In general, stellar dispersions are higher in the thick disc and inner regions, and lower in the thin disc and dust lane. Both positions show a negative velocity dispersion gradient with projected radius: velocity dispersions in the inner region are higher than in the outer disc regions, as expected. In the mean, stellar velocity dispersions in pos A and pos B are essentially the same (82\,$\rm km \, s^{-1}$).

\begin{figure}
    \centering
    \includegraphics[width=1.0\linewidth]{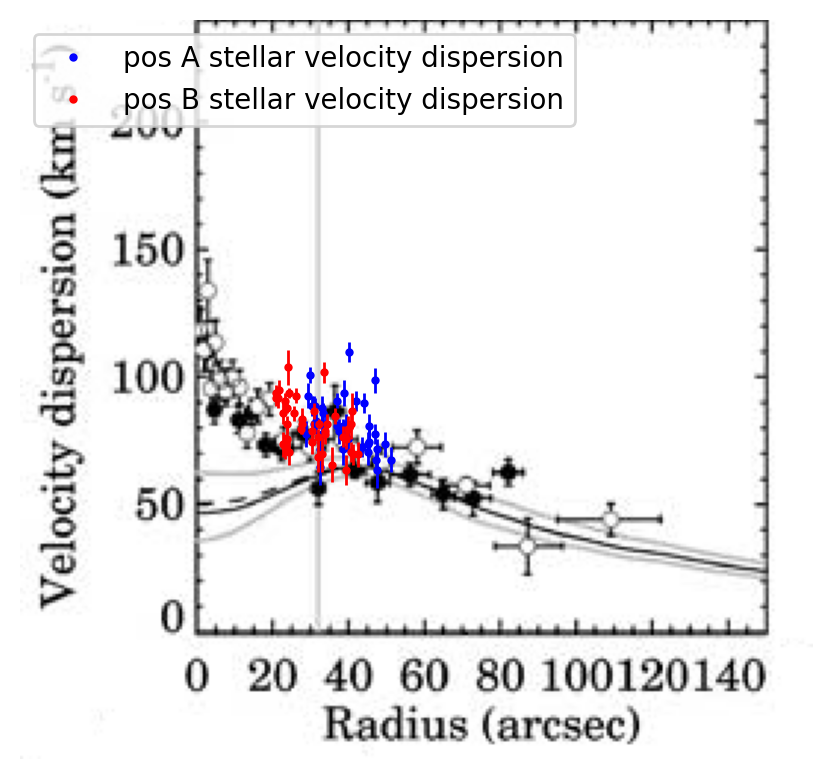}
        \caption{Plot of our stellar velocity dispersions over the stellar velocity dispersion curve from \citet{2005MNRAS.358..481K}, errors in this plot are \textsc{ppxf} errors. We thank Prof. Piet van der Kruit for allowing us to use this image in our paper.}
    \label{fig:rdvdb}
\end{figure}

We plot our stellar velocity dispersions over the stellar velocity dispersion curve from \citet{2005MNRAS.358..481K} in Figure~\ref{fig:rdvdb}. In comparison, our velocity dispersions generally follow the trend of their curve well: those significantly higher than the curve are mainly bins at high $z$ (their data comes from a single long slit spectrum and low height). Their curve shows jumps in velocity dispersion on both sides of the galaxy, at a projected radius about 35" and about 5" south of the mid-plane (see also \citet{2002ASPC..275...47K}), which are also seen in our bins A21-23t22-25 and B16-19t23-30: amplitudes of these jumps are about 10\,$\rm km \, s^{-1}$ in velocity dispersion but without any corresponding feature in stellar velocities. Combined with the 20° bar angle in IC 2531 reported by \citet{2010ASPC..424..261A}, these jumps in the velocity dispersion are consistent with the secondary peak in the simulated intermediate-bar edge-on galaxy’s velocity dispersion profile of~\citet{2005ApJ...626..159B}.

Stellar velocity dispersion in a galaxy disc is proportional to $e^{-r/2h_r}$ \citep{2011ARA&A..49..301V}. The scale length ${h_r}$ of IC 2531’s thin disc is about 45" \citep{2016A&A...592A..71M}. As the width of our fields is 25" in radius, the ratio between stellar velocity dispersion at the inner and outer edges of our fields would be about 1.31. We compare this ratio with observed ratios in our thin disc low $z$ fields, for pos A south we compare the second innermost bins as the innermost bin’s dispersion is abnormally low, maybe due to projection. For other fields we use the innermost and outermost bins. For pos A north and south, pos B north and south, observed ratios are 1.26, 1.31, 1.34, 1.31. This shows our velocity dispersion gradients agree well with what we know about galaxy discs and those high dispersions at small radii are not due to bulge contamination.

\begin{figure}
    \centering
    \includegraphics[width=1.0\linewidth]{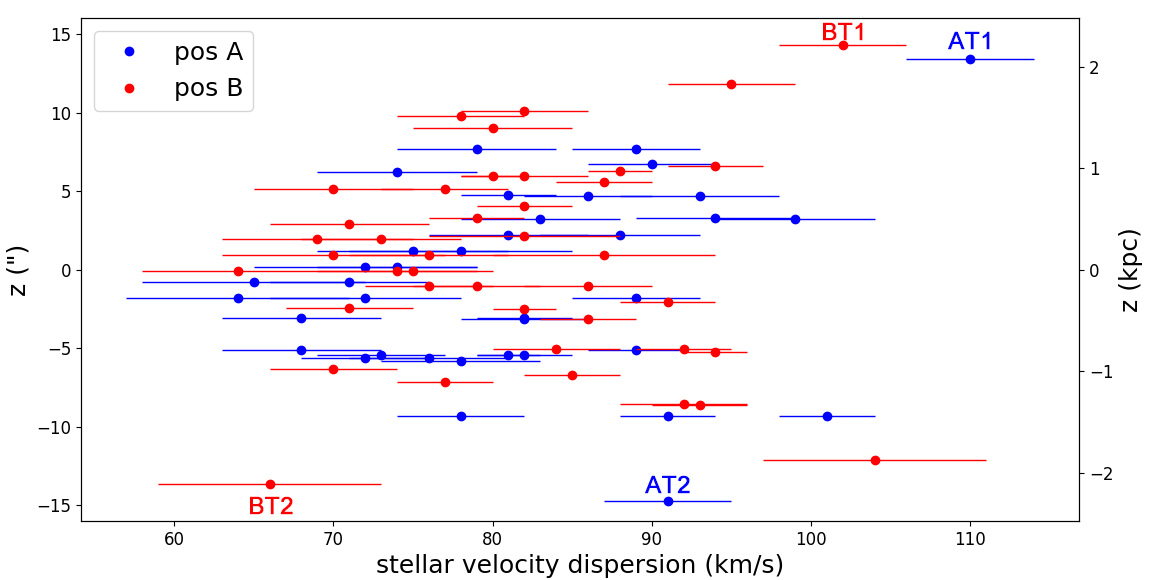}
        \caption{Plot of our stellar velocity dispersions vs $z$, errors in this plot are \textsc{ppxf} errors.}
    \label{fig:dv_xAB}
\end{figure}

To show the vertical variation of velocity dispersions, we plot stellar velocity dispersions vs height in Figure~\ref{fig:dv_xAB}. It is obvious that stellar velocity dispersions gradually increase with height, with an exception of BT2 which is at the bottom left corner of the figure.

The velocity dispersion of BT2 is significantly lower than those in the other three thick disc bins, and is one of the lowest among all bins. There are no obvious issues with the other parameters in this bin: the mean velocity is equal to that of BT1, and the stellar population of BT2 is old, metal-poor and alpha-enhanced as expected. Several other tests were performed to check for this anomalous result: we tried different sub-binning regions in BT2, result shows majority of BT2’s region have a low dispersion of around 63\,km\,s$^{-1}$. We tried Fourier power spectrum analysis of BT2’s unsmoothed spectra, result shows it has a dispersion of about 52\,km\,s$^{-1}$. All those additional analyses suggest the low dispersion of BT2 is likely to be true, but we still do not understand why its velocity dispersion is so low.

\subsubsection{Ionised gas velocity}

\begin{figure*}
    \centering
    \includegraphics[width=0.9\linewidth]{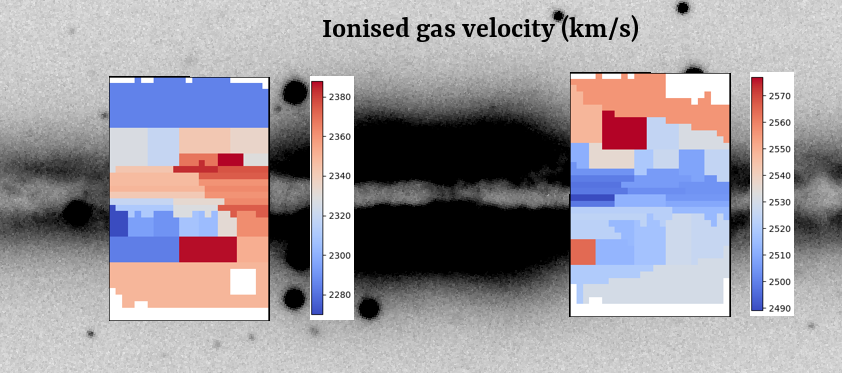}
        \caption{Similar to Figure~\ref{fig:ke}, but for ionised gas velocity fields.}
    \label{fig:gv}
\end{figure*}

We show the map of ionised gas velocities in Figure~\ref{fig:gv}. Ionised gas velocities are more uniform in pos B compared to those in pos A, and are generally uniform in the dust lane. 

In pos B, the most extreme (highest) ionised gas velocity ($>2550 \,\rm km \, s^{-1}$) appears in the inner north part of the thin disc and BT1. B1-4t27-30 is the only bin in the southern half with a similarly high ionised gas velocity, and shows a jump of about 40\,$\rm km \, s^{-1}$ compared to surrounding bins. In other thin disc bins and in BT2, the ionised gas velocities are generally around 2520\,$\rm km \, s^{-1}$, while the dust lane has the lowest ionised gas velocity (about 2510\,$\rm km \, s^{-1}$), as we might expect due to projection effects.

In pos A, the most extreme (lowest) ionised gas velocity appears in AT1 and the middle to outer parts in the southern thin disc bin, except for A12-20t26-29 which has one of the highest ionised gas velocities in pos A, and shows a jump of 20-100\,$\rm km \, s^{-1}$ compared to surrounding bins. Another area with high ionised gas velocity is the inner top part of the dust lane and the adjacent thin disc bins. It seems likely that these bins lie at relatively large actual radii along lines-of-sight toward us.

Ionised gas (from H\,\textsc{ii} regions) velocities are potentially useful for interpreting the depths of our lines-of-sight, as H\,\textsc{ii} regions are likely to follow the kinematics of H\,\textsc{i} near the galactic plane. By comparing ionised gas velocities with the H\,\textsc{i} $(p,v)$ diagram, we can get an idea of how deep the ionised gas lies along the line-of-sight in each bin by comparing the ionised gas velocities with those of the H\,\textsc{i} tangent point velocities. Tangent points (i.e., closest approach of the line-of-sight to the galactic centre, where the velocity due to the rotation field is the largest) appear near the extreme (high in pos B, low in pos A) velocity edges of H\,\textsc{i} distributions in the $(p,v)$ diagram.

\begin{figure}
    \centering
    \begin{minipage}{0.99\linewidth}
        \centering
        \includegraphics[width=1\linewidth]{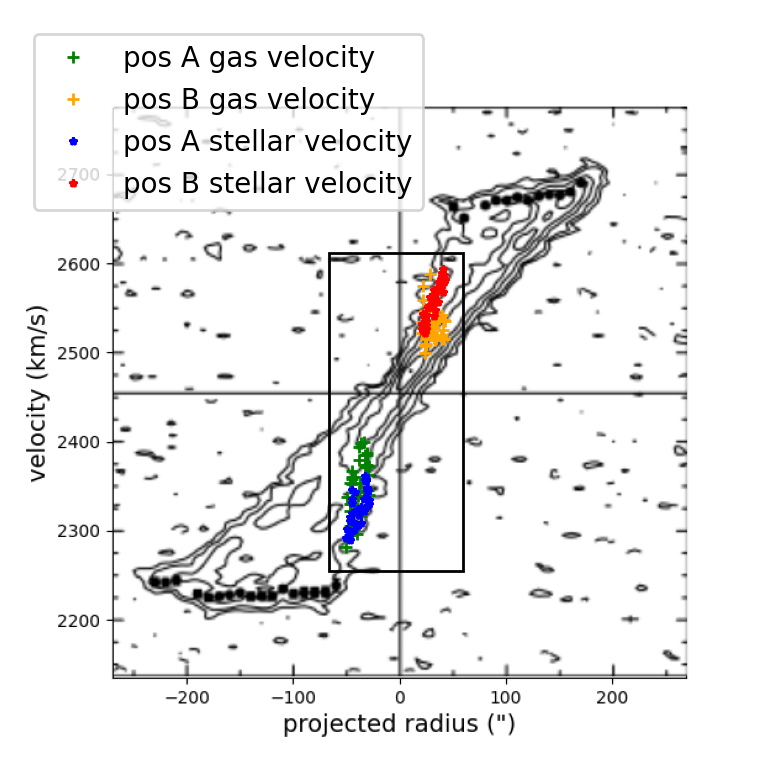}
    \end{minipage}
    \begin{minipage}{0.8\linewidth}
        \centering
        \includegraphics[width=1\linewidth]{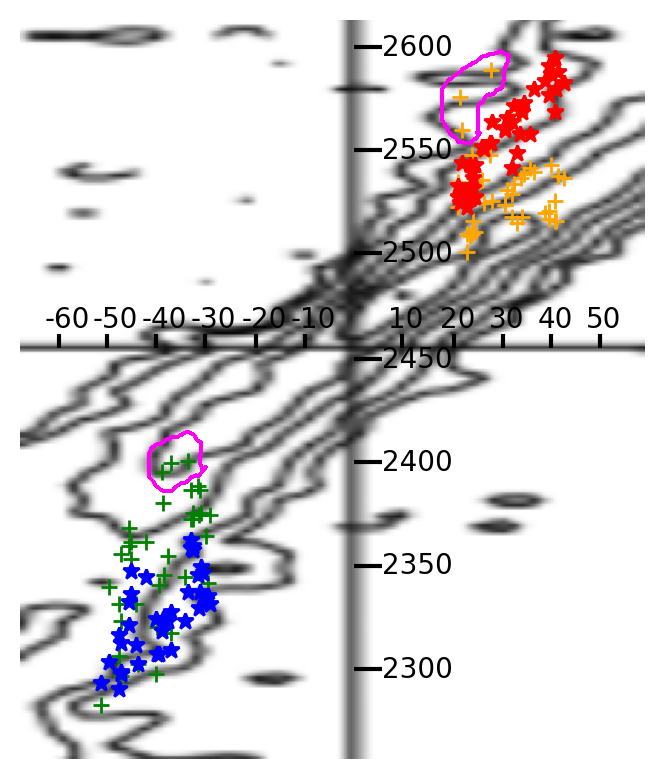}
    \end{minipage}
    \caption{
    Top panel: all our optical line-of-sight velocities (stellar + gas) $(p,v)$ data plotted over the H\,\textsc{i} $(p,v)$ data from~\citet{2004MNRAS.352..768K}. Tangent point velocities in these figures appear near extreme velocity (high in pos B, low in pos A) edges of H\,\textsc{i} velocities. We thank Prof. Piet van der Kruit for allowing us to use this image in our paper.
    Bottom panel: A zoomed in, horizontal stretched (for clarity) distribution of the black square in the top panel. Closed purple curves identify optical gas velocities in each position that distribute differently from most other bins.
    }
  \label{fig:gs2kz}
\end{figure}

We plot $(p,v)$ data points (adjusted to \citet{2004MNRAS.352..768K}’s systemic velocity) of our stellar and ionised gas on the H\,\textsc{i} $(p,v)$ diagram from~\citet{2004MNRAS.352..768K} in Figure~\ref{fig:gs2kz}. In both pos A and pos B, the dust lane bins have ionised gas velocities far away from tangent point velocities, where our effective optical lines of sight do not reach into the tangent point regions. For ionized gas, in pos A, most of the other lines of sight appear to be close to tangent points, except for A18-21t13-14, A11-17d14-15 and A12-20t26-29. They appear red in Figure~\ref{fig:gv} and are identified with a closed purple curve in Figure~\ref{fig:gs2kz}. On the other hand, in pos B, Figure~\ref{fig:gs2kz} shows that most of our optical lines of sight are far away from tangent point velocities, with exceptions of bin B1-5t3-11, B6-12t7-12 and B1-4t27-30 (identified with closed purple curves in Figure~\ref{fig:gs2kz} as well). They are likely to be located at relatively small actual radii, and appear orange and red in Figure~\ref{fig:gv}. 

The stellar velocities in pos B are close to the tangent point velocities, and distinct from the ionised gas velocities at large projected radii. This suggests that most of the stars we observed in pos B lie close to the tangent points, but this is not the case for ionised gas. The stellar velocities in pos A are also closer to tangent point velocities and are slightly separate from ionised gas velocities at small projected radii. These results suggest that the ionised gas velocities may be associated with H\,\textsc{ii} regions and star formation regions that are not located near tangent points, perhaps instead in spiral arms along the lines-of-sight.

In both positions, Figure~\ref{fig:gs2kz} bottom panel shows a concentration of stellar and ionised gas velocity points in each field (e.g. at $R=-45$") that extend over a range of velocity. These come from bins (especially dust bins) in each field with a similar projected radius range (cf. Table~\ref{tab:dpj}). For these bins, although their velocities are plotted at roughly the same mean projected radius, they are affected by dust attenuation differently, and therefore have different actual radii and different velocities.

From Figure~\ref{fig:gs2kz}, there is a marked difference between the H\,\textsc{i} $(p,v)$ diagram for east (negative projected radius) and west (positive projected radius) sides: the extent of the $(p,v)$ diagram covered on the east side is much larger than on the west side. Also, the range of H\,\textsc{i} and optical velocities are rather similar in pos B, but H\,\textsc{i} velocities have a broader velocity range than the optical velocities in pos A. It seems likely that the distribution of H\,\textsc{i} is more extended along the line-of-sight on the east side of IC 2531 than the west side.

\begin{figure*}
    \centering
    \includegraphics[width=1.0\linewidth]{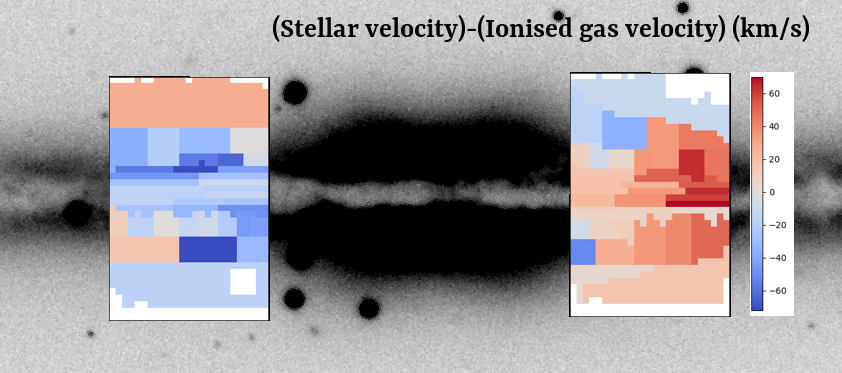}
        \caption{Plot of differences between stellar and ionised gas velocity $(v_{star}$-$v_{gas})$. Blue regions in pos A and red regions in pos B are where the stellar velocity is closer to the tangent point velocity, and stars are closer to tangent points in the mean.}
  \label{fig:ABvs-vg}
\end{figure*}

To understand which component is located closer to the tangent points, in Figure~\ref{fig:ABvs-vg} we show a map of the differences between the stellar and ionised gas velocities. For pos A, in a blue region, the stellar velocity is closer to the tangent point velocity, and the stars lie closer to tangent points in the mean; in a red region, the ionised gas velocity is closer to the tangent point velocity, and the ionised gas is closer to tangent points in the mean. The correlation is opposite for pos B. Figure~\ref{fig:ABvs-vg} shows that in most bins, stars are closer to the tangent points, and the ionised gas is at relatively large actual radii compared to the stellar component. Exceptions are usually caused by extreme ionised gas velocities (rather than stellar velocities, cf. Figures~\ref{fig:gs2kz}), where the ionised gas is at relatively small radii and closer to tangent points than the stars. One interesting phenomenon is that the northern and southern thick disc bins are exactly opposite in Figure~\ref{fig:ABvs-vg}, suggesting there might be a structural difference between the northern and southern halves of IC 2531: in the northern thick disc bins, the ionised gas is closer to tangent points, whereas in the southern thick disc bins, the stars are closer to tangent points.

\subsection{Stellar Population}
\label{sec:3.2}
We combined spectra from all the dust region bins, thin disc bins and thick disc bins respectively according to their classification (and position), and created five new large bins in each position, namely, the total dust region bin, total thin disc bin, the north or south thin disc bins, and the total thick disc bin, and got \textsc{ppxf} fitting results for their spectra as the light weighted average stellar parameter of different projected regions in IC 2531, those results are shown in Table~\ref{tab:avg vl}. These large bins and their fitting results are not included in the figures in this paper. Comparing [M/H] and [$\alpha$/Fe] in Table~\ref{tab:avg vl} with Figure 4 of \citep{2015ApJ...808..132H} shows, stellar populations in different regions of IC 2531 are most similar to the following regions of the Milky Way: 

Pos B dust region and both north and south thin disc bins, pos A south thin disc and AT2 are most similar to \citep{2015ApJ...808..132H}’s $0.5<\lvert z \rvert<1$ kpc, $5<R<7$ kpc; 

Pos A dust region is most similar to $0.5<\lvert z \rvert<1$ kpc, $9<R<11$ kpc;

Pos A north thin disc bin is most similar to $0<\lvert z \rvert<0.5$ kpc, $9<R<11$ kpc; 

AT1 is most similar to $0<\lvert z \rvert<0.5$ kpc, $5<R<7$ kpc; 

BT1 is most similar to $1<\lvert z \rvert<2$ kpc, $9<R<11$ kpc; 

BT2 is most similar to $1<\lvert z \rvert<2$ kpc, $3<R<5$ kpc. 

\begin{table*}
  \centering
  \caption{IC 2531's light weighted average stellar population parameters and ${\rm H}\upbeta$ emission line peak intensities (relative flux to units of counts, cf. Figure~\ref{fig:ftr}) in different projected regions. On average, IC 2531's thin and thick discs show similar stellar population characteristics to the Milky Way, with the exception of pos A's north thick disc bin AT1, which has all characteristics of a thin disc.}
    \begin{tabular}{|c|c|c|c|c|c|c|c|c|}
    \hline
          &       & dust region & \multicolumn{3}{c|}{thin disc} & \multicolumn{3}{c|}{thick disc} \\
    \hline
    \multirow{4}[2]{*}{pos A} &       &       & \multicolumn{1}{c}{total thin} & \multicolumn{1}{c}{north} & south & \multicolumn{1}{c}{total thick} & \multicolumn{1}{c}{AT1} & AT2 \\
          & lg (age) & 0.65±0.10 & \multicolumn{1}{c}{0.82±0.06} & \multicolumn{1}{c}{0.91±0.06} & 0.78±0.06 & \multicolumn{1}{c}{0.64±0.14} & \multicolumn{1}{c}{0.59±0.18} & 0.94±0.12 \\
          & [M/H] & -0.09±0.08 & \multicolumn{1}{c}{-0.10±0.05} & \multicolumn{1}{c}{-0.14±0.05} & -0.12±0.06 & \multicolumn{1}{c}{0.14±0.13} & \multicolumn{1}{c}{0.37±0.08} & -0.25±0.10 \\
          & [$\alpha$/Fe] & 0.03±0.03 & \multicolumn{1}{c}{0.09±0.04} & \multicolumn{1}{c}{0.05±0.03} & 0.14±0.03 & \multicolumn{1}{c}{0.14±0.06} & \multicolumn{1}{c}{0.03±0.05} & 0.16±0.07 \\
          & ${\rm H}\upbeta$ & 0.62 & \multicolumn{1}{c}{0.25} & \multicolumn{1}{c}{0.26} & 0.25 & \multicolumn{1}{c}{0.32} & \multicolumn{1}{c}{0.17} & 0.42 \\

    \multirow{4}[2]{*}{pos B} &       &       & \multicolumn{1}{c}{total} & \multicolumn{1}{c}{north} & south & \multicolumn{1}{c}{total} & \multicolumn{1}{c}{BT1} & BT2 \\
          & lg (age) & 0.82±0.04 & \multicolumn{1}{c}{0.95±0.03} & \multicolumn{1}{c}{0.98±0.03} & 0.95±0.04 & \multicolumn{1}{c}{1.12±0.04} & \multicolumn{1}{c}{1.07±0.07} & 1.12±0.09 \\
          & [M/H] & -0.13±0.04 & \multicolumn{1}{c}{-0.13±0.03} & \multicolumn{1}{c}{-0.14±0.03} & -0.13±0.04 & \multicolumn{1}{c}{-0.27±0.05} & \multicolumn{1}{c}{-0.38±0.06} & -0.24±0.12 \\
          & [$\alpha$/Fe] & 0.15±0.02 & \multicolumn{1}{c}{0.20±0.03} & \multicolumn{1}{c}{0.13±0.02} & 0.21±0.03 & \multicolumn{1}{c}{0.26±0.04} & \multicolumn{1}{c}{0.15±0.05} & 0.31±0.03 \\
          & ${\rm H}\upbeta$ & 1.13 & \multicolumn{1}{c}{0.46} & \multicolumn{1}{c}{0.47} & 0.45 & \multicolumn{1}{c}{0.49} & \multicolumn{1}{c}{0.37} & 0.64 \\
    \hline
    \end{tabular}%
  \label{tab:avg vl}%
\end{table*}

Table~\ref{tab:avg vl} shows in pos A southern half and pos B, the average stellar populations for discs above the dust plane generally have stellar populations lying between the Milky Way’s chemical thin and thick disc (this agrees with our analysis of what part of the disc we are looking at in section~\ref{sec:2.4}). The thick disc is older, more metal-poor, but similarly alpha-enhanced relative to its corresponding thin disc. However, one of thick disc bins (BT2) clearly shows as an alpha-rich chemical thick disc.

An interesting finding from this comparison is that pos A northern half (north thin disc and AT1) have stellar populations most similar to the Milky Way’s low $z$ ($0<\lvert z \rvert<0.5$ kpc), while most of the other regions we see in IC 2531 are most similar to the Milky Way’s medium $z$ ($0.5<\lvert z \rvert<1$ kpc). Pos B thick disc bins are most similar to the Milky Way’s high $z$ ($1<\lvert z \rvert<2$ kpc) as expected.

Compared to kinematics, stellar populations are generally harder to estimate and come with larger uncertainties. An empirical upper limit on uncertainties in derived parameters can be independently estimated from the scatter of parameters, e.g.~\citet{2016A&A...593L...6C}. For example, the [M/H] distribution in thin discs is expected to show a negative gradient with radius; the observed scatter about such a gradient therefore gives a measure of uncertainties in our derived [M/H] values.

\begin{figure}
    \centering
    \begin{minipage}{0.99\linewidth}
        \centering
        \includegraphics[width=1\linewidth]{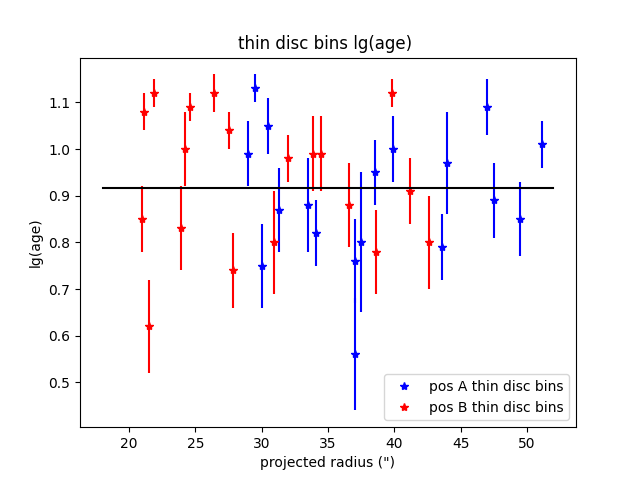}
    \end{minipage}
    \begin{minipage}{0.99\linewidth}
        \centering
        \includegraphics[width=1\linewidth]{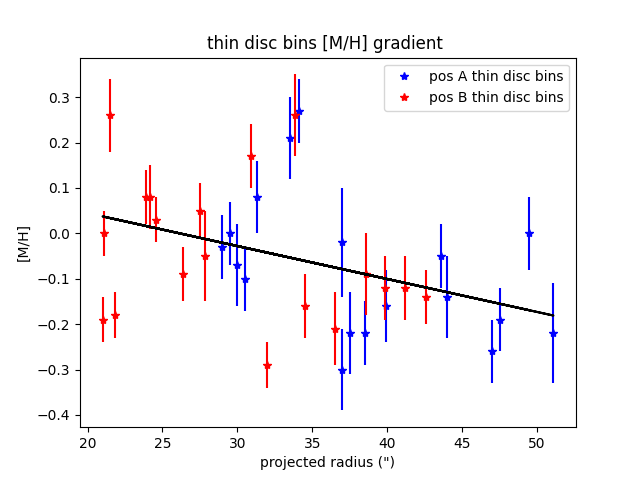}
    \end{minipage}    
    \begin{minipage}{0.99\linewidth}
        \centering
        \includegraphics[width=1\linewidth]{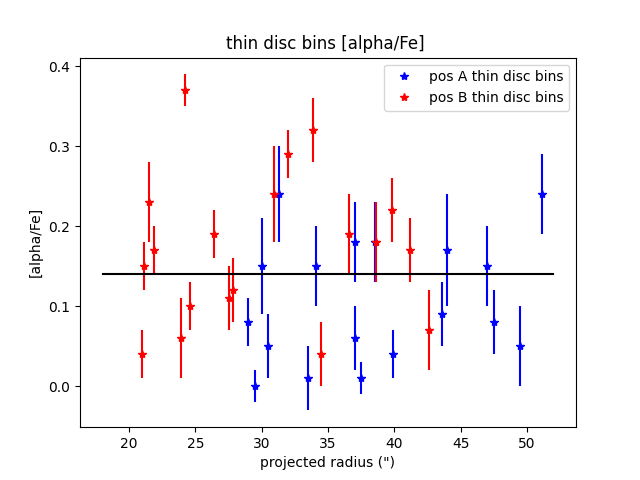}
    \end{minipage}    
    \caption{Stellar population in thin discs used to estimate empirical upper limits on uncertainties in our \textsc{ppxf} estimation, the black line is the fitting line. Errors in these plots are Monte Carlo simulations’ 1$\sigma$ uncertainties. Top panel: age, we assume on average it is a constant in the thin disc as there are no clear trends. Middle panel: [M/H] , like in the Milky Way, there is an obvious negative gradients with radius. Bottom panel: [$\alpha$/Fe], similar as age, assuming on average it is a constant in the thin disc.}
  \label{fig:fit}
\end{figure}

To derive empirical upper limits on uncertainties, we use all the thin disc bins. The negative [M/H] gradient is clearly visible in Figure~\ref{fig:fit}, with a calculated gradient about $-0.046$ dex/kpc, compare to about $-0.06$ dex/kpc in the Milky Way \citep{2006AJ....132..902L}. There are no clear trends in age and [$\alpha$/Fe] so we adopt constant age and [$\alpha$/Fe]. With these assumptions, derived upper limits of uncertainties in lg(age), [M/H], and [$\alpha$/Fe] are approximately 0.14, 0.14, and 0.09 respectively. These uncertainties are in good agreement with those yielded by our Monte Carlo simulations. For thin disc bins, calculated covariances in lg(age) and [M/H], lg(age) and [$\alpha$/Fe], and [M/H] and [$\alpha$/Fe] are about $-0.005$, 0.001, 0.002 respectively, and are typically an order of magnitude smaller than the variance terms, which shows there is no strong relation between stellar population parameters in the thin disc of IC 2531 and our spectra fitting.

Monte Carlo simulations of the \textsc{ppxf} fitting process are an alternate way to estimate uncertainties, e.g.~\citet{2021MNRAS.508.2458M}. In each bin, we created 100 mock spectra by adding Gaussian random noise (where the amplitude is derived from the \textsc{ppxf} fitting residual, e.g. the green curve in Figure~\ref{fig:ftr}) to the best stellar fit returned by \textsc{ppxf}. We then ran these mock spectra through \textsc{ppxf} to obtain parameter distributions. Finally, we calculated 1$\sigma$ standard deviations for each parameter as their uncertainties.

Overall, as shown in Tables~\ref{tab:MCA} and~\ref{tab:MCB}, the uncertainties of our \textsc{ppxf} fitting results are satisfactory: most of the uncertainties are around or under 0.1. In pos A, average uncertainties in lg(age), [M/H], and [$\alpha$/Fe] are about 0.12, 0.09, and 0.05 respectively, and in pos B they are about 0.07, 0.08, and 0.04.

\subsubsection{Age}

\begin{figure*}
    \centering
    \includegraphics[width=0.9\linewidth]{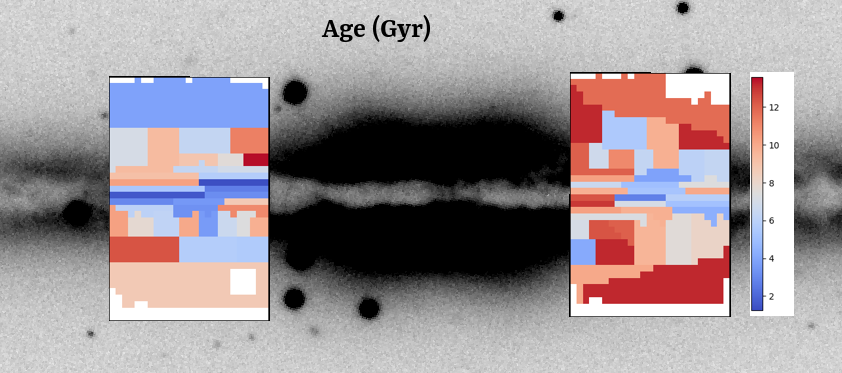}
        \caption{Maps of the stellar age for pos A and B.}
    \label{fig:po}
\end{figure*}

\begin{figure}
    \centering
    \includegraphics[width=1.0\linewidth]{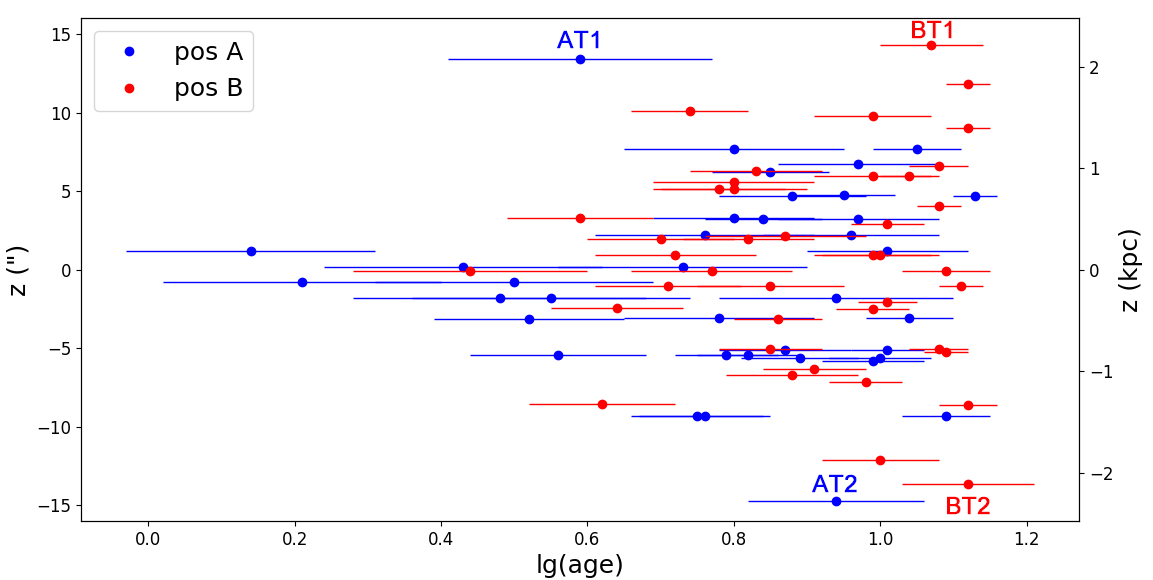}
        \caption{Similar to Figure~\ref{fig:dv_xAB}, but for stellar age, errors in this plot are Monte Carlo simulations’ 1$\sigma$ uncertainties.}
    \label{fig:age_xAB}
\end{figure}

According to Table~\ref{tab:avg vl}, the age map in Figure~\ref{fig:po} and age vs height in Figure~\ref{fig:age_xAB}, in general, pos B exhibits older stellar populations than pos A. Although old stars can be found at every galactic height, younger and younger stars appear as the height drops toward the central plane, this results in a positive age gradient with height for both positions: on average, thick disc regions are the oldest (except for AT1 on the top center of Figure~\ref{fig:age_xAB}, shows as an outlier, it has a young age that otherwise can only be found at a low height), and the dust lane is the youngest, while the thin disc regions are in between. Pos B also shows a generally old region at a small projected radius.
 
Pos B shows an obvious positive age gradient along the vertical axis, with mean age becoming older at higher $z$. One can easily identify a very old thick disc component in Figure~\ref{fig:po}: the two thick disc bins BT1 and BT2 are both about 12 Gyr old. The small projected radius ($x<10$) of pos B is generally old, with most bins older than 10 Gyr, even dust lane bins in this region tend to be old ($>10$ Gyr). The dust lane outside that is the youngest region in pos B (mostly 3-7 Gyr). Ages of the thin disc bins generally lie between those values, suggesting that the thin disc and dust lane have extended star formation histories compared to the thick disc. 

As shown in Figure~\ref{fig:po}, pos A is younger than pos B. There is also a hint of a positive vertical age gradient in the southern half of pos A, but not as obvious as in pos B. Stars in the dust lane are also young (mostly $<6$ Gyr); the youngest bins (A15-25d17 and A1-15d19) are only about 1.5 Gyr old. Thin disc bins show a mix of middle (6-7 Gyr) and old ($>9$ Gyr) populations. The age of the thick disc region AT2 is about 9 Gyr. The oldest stars appear in a bin (A22-25t13-14, about 13 Gyr) with low $z$ and relatively small projected radius.

Despite the old stellar populations, low-level star formation is present throughout the disc as shown by the existence of H\,\textsc{ii} regions indicated by ${\rm H}\upbeta$ emission line which appears in most bins’ spectra (e.g. see Figure~\ref{fig:ftr}), especially in dust lane bins.

\subsubsection{Metallicity}

\begin{figure*}
    \centering
    \includegraphics[width=0.9\linewidth]{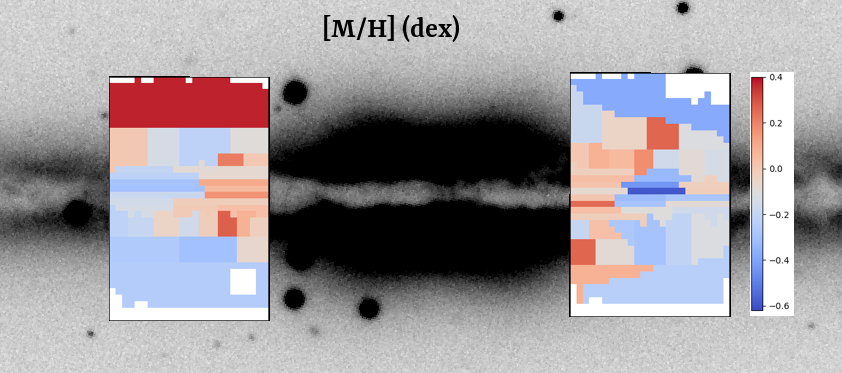}
        \caption{Maps of [M/H] for pos A and B.}
    \label{fig:metal}
\end{figure*}

\begin{figure}
    \centering
    \includegraphics[width=1.0\linewidth]{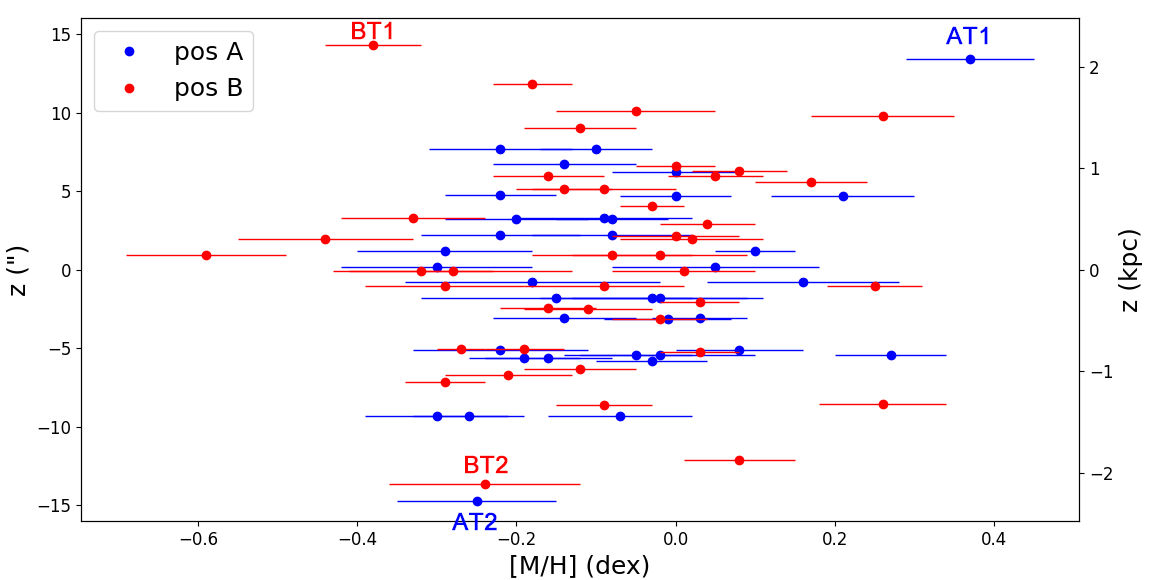}
        \caption{Similar to Figure~\ref{fig:age_xAB}, but for [M/H].}
    \label{fig:metal_xAB}
\end{figure}

According to Table~\ref{tab:avg vl}, the [M/H] map in Figure~\ref{fig:metal} and [M/H] vs height in Figure~\ref{fig:metal_xAB}, with the exception of AT1 (again shows as an outlier on the top right corner of Figure~\ref{fig:metal_xAB}), IC 2531's thick disc has lower metallicity compared to thin disc in general. There is a negative radial metallicity gradient with a [M/H] jump in the thin disc, resulting in a large [M/H] range from about $-0.3$ to 0.3, and the inner thin disc (including the dust lane) is the most metal-rich region. Some bins in the dust lane have a very low [M/H], probably due to stars at large actual radii. 

As shown in the Figure~\ref{fig:fit} middle panel, IC 2531’s thin disc shows an obvious negative radial [M/H] gradient with a jump of [M/H] to about 0.25 at a projected radius of about 34" (5.1 kpc) on both sides, and based on our fitted gradient in [M/H] ($-0.046$ dex/kpc) for its entire population, the calculated [M/H] at a projected radius of 14 kpc is about $-0.45$. In comparison, the peak of the Milky Way's [Fe/H] is about 0.3, occurring within a radius between 3 to 7 kpc and $0<\lvert z \rvert<0.6$ kpc \citep{2015ApJ...808..132H, 2021A&A...653A.143W}, it also has a gradient of $-0.06$ dex/kpc \citep{2006AJ....132..902L} and the [Fe/H] is about $-0.48$ between radius from 13 to 15 kpc \citep{2015ApJ...808..132H}.

In Figure~\ref{fig:metal}, pos B shows a low metallicity thick disc. BT1 has a metallicity of $-0.38$; the metallicity of BT2 is $-0.24$, but it is still lower than that of any other bin in the southern half of pos B, except it is slightly higher than B11-15t23-30. In the middle of the dust lane, there is a very metal-poor region ([M/H]$<-0.3$), with the lowest reaching $-0.59$ (B10-18d19). This may be due to these bins contain thick disc stars or thin disc stars with large orbital radii that are in front of the dust. Pos B has a relatively high [M/H] at small projected radius ($x<10$), as the result of the radial gradient, where most bins have [M/H]$\geq$0. The [M/H] jump discussed above can be seen on pos B northern half as another metal-rich region (B13-17t8-12 and B11-13t13-16). However, this feature is missing in pos B southern half. Other thin disc bins in pos B are relatively metal-poor ([M/H]$<-0.1$).

As can be seen in Figure~\ref{fig:metal}, pos A has a low metallicity thick disc with a metallicity of $-0.25$ on the south side (AT2). The metallicity distribution in the thin disc and dust lane in pos A are essentially symmetrical in $z$. Because of the radial [M/H] gradient, pos A bins with small projected radius and low $z$ are also relatively metal-rich ([M/H]$>-0.1$). In pos A, the [M/H] jump appears at both northern and southern half of (A18-21t13-14 and A18-20t22-25), these two bins have the highest metallicity in pos A thin disc region (0.21 and 0.27). In the dust lane’s outer part ($x$ from 1 to 15), where stellar velocities decrease with height, metallicity also increases with height. This is another sign that we are looking into deeper parts of the galaxy as our line-of-sight moves higher above the central plane. In the southern half of pos A, the distribution of [M/H] is quite uniform (about $-0.26$) for $y>26$, except A21-25t26-29 which has the smallest project radius. The metallicity distribution in the northern half of the thin disc region in pos A is similar to the southern half, except it increases to 0 at A1-6t9-15.

\subsubsection{[$\alpha$/Fe]}

\begin{figure*}
    \centering
    \includegraphics[width=0.9\linewidth]{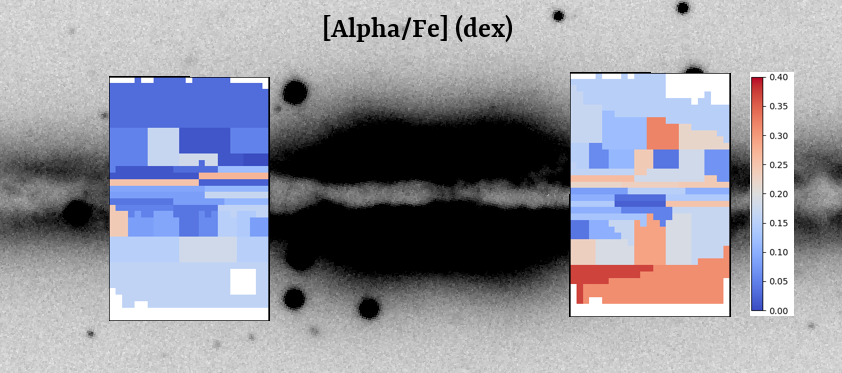}
        \caption{Maps of [$\alpha$/Fe] for pos A and B.}
    \label{fig:alpha}
\end{figure*}

\begin{figure}
    \centering
    \includegraphics[width=1.0\linewidth]{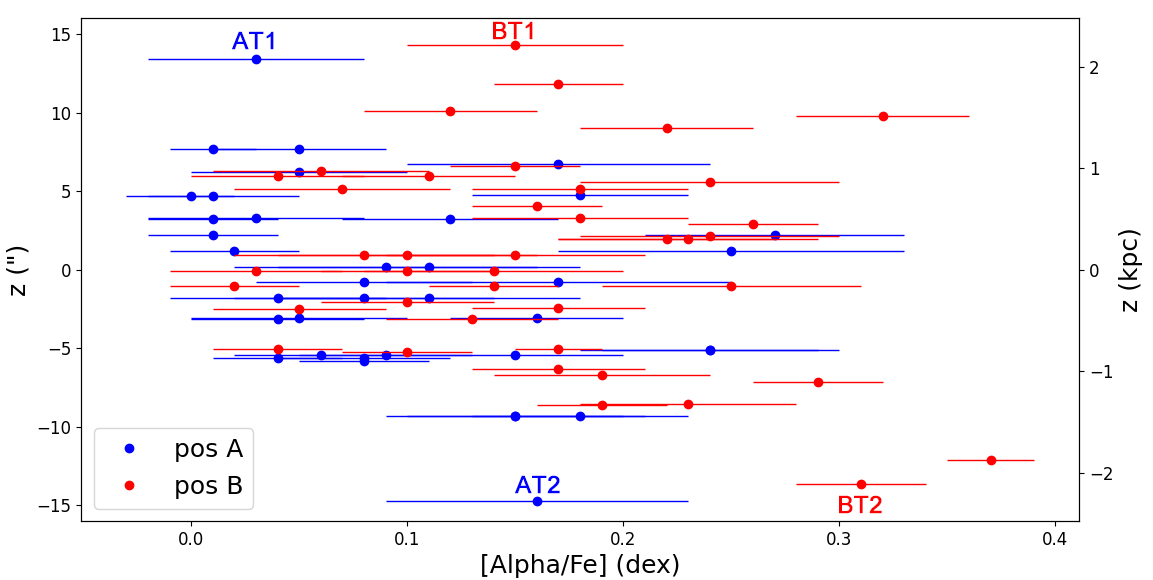}
        \caption{Similar to Figure~\ref{fig:age_xAB}, but for [$\alpha$/Fe].}
    \label{fig:alpha_xAB}
\end{figure}

According to Table~\ref{tab:avg vl}, the [$\alpha$/Fe] map in Figure~\ref{fig:alpha} and [$\alpha$/Fe] vs height in Figure~\ref {fig:alpha_xAB}, except for AT1 (located at the top left corner of Figure~\ref {fig:alpha_xAB}), bins with $\lvert z \rvert>9$ generally have enhanced alpha ([$\alpha$/Fe]$>0.1$), high alpha stars ([$\alpha$/Fe]$>0.3$) only exists at high $z$. Compared to their corresponding thin disc (cf. Table~\ref{tab:avg vl}), two of the thick disc bins (AT2 and BT1) have similar alpha enhancement, one thick disc bin (BT2) is clearly alpha-rich. The southern half of the galaxy appears more alpha-rich than the northern half. Bins with $y \approx 17$ also show an alpha enhancement, which is a possible indication that IC 2531’s chemical thick disc extends further radially than the chemical thin disc and dust. 

Figure~\ref{fig:alpha} shows in pos B the most alpha-rich stars appear in high $z$ and shows a chemical thick disc on the south side. Bin B1-13t31-36's [$\alpha$/Fe] reaches 0.37, which is the most alpha-rich bin in our observations, and is close to the most alpha-rich stars with [$\alpha$/Fe] $=0.4$ in the Milky Way's thick disc. BT2 also has a high [$\alpha$/Fe] value (0.31). The [$\alpha$/Fe] of BT1 is only 0.15, which is similar to the average [$\alpha$/Fe] in the northern thin disc of pos B. But BT1 may also be affected by the same warp as AT1 (see Section~\ref{sec:3.3}), where further analysis shows that BT1 has an [$\alpha$/Fe] of about 0.23 after excluding the top four-pixel rows. In this case, its [$\alpha$/Fe] would be higher than all except for B13-17t8-12 (0.32) and B11-13t13-16 (0.24, basically the same considering the uncertainties). These two bins plus B18-25t9-12 form an alpha-rich region in the thin disc, and a similar structure (B11-15t23-30) appears symmetrically in the southern half. The [$\alpha$/Fe] of other thin disc bins are generally less than 0.2. Although the dust lane is generally the most alpha-poor region, especially in the lower middle part where [$\alpha$/Fe]$<0.05$, the upper dust lane in pos B is still relatively alpha-rich ($>0.2$).

In general, pos B is more alpha-rich than pos A. As shown in Figure~\ref{fig:alpha}, in the southern half of pos A, [$\alpha$/Fe] is relatively uniform (about 0.16) for $y>26$, similar to the observed trend in metallicity. Thin disc bins closer to the dust lane are alpha-poor by comparison, with [$\alpha$/Fe] generally around 0.1, except A1-3t21-25 which has [$\alpha$/Fe] $=0.24$. A7-11t9-14 and A12-17t13-14 show a significant [$\alpha$/Fe] increase: their [$\alpha$/Fe] are about 0.17, which is higher by more than 0.1 in comparison with other thin disc bins in pos A northern half. In pos A, [$\alpha$/Fe] changes dramatically in the dust lane's upper part, from nearly zero to about 0.26. [$\alpha$/Fe] values are more uniform in the dust lane region below that, with [$\alpha$/Fe] of about 0.14 for the inner half and about 0.06 for the outer half.

For both positions, the most alpha-poor bins appear in and around the dust lane with almost zero [$\alpha$/Fe]. Here, it is likely that we are seeing the thin disc through some dust. However, some bins in the dust lane (especially at $y \approx 17$) show an alpha enhancement (about 0.25), as a horizontal orange bar in the upper dust lane in Figure~\ref{fig:alpha}, which may be caused by the chemical thick disc extending further radially than the chemical thin disc and dust, and wrapping around the chemical thin disc. If this is the case, the spatial distribution of the chemical thick disc in IC 2531 would be different from the Milky Way’s chemical thick disc; as~\citet{2015ApJ...808..132H} shows, there are basically no alpha-rich stars at large radii near the central plane. This agrees with \citet{2016A&A...592A..71M}’s IC 2531’s radial profile model in their Fig 3, which shows the thick disc of IC 2531 has larger scale length, extends further than the thin disc and contributes more in overall light profile than thin disc at about projected radius $R>130$".

To summarise, pos B appears older and more alpha-rich than pos A. With the exception of bin AT1, on average, the thick disc bins in IC 2531 appear older, more metal-poor and have a similarly enhanced [$\alpha$/Fe] compared to the thin disc bins, which is generally consistent with what has been observed in the Milky Way.

\subsection{AT1 discussion}
\label{sec:3.3}

\subsubsection{Observed effects of the warp}

Bin AT1 has unexpected properties. Although its position in projection appears to be a thick disc region, it has all the characteristics of a thin disc's stellar population: young (3.9 Gyr), metal-rich (0.37), and alpha-poor (0.03). However, it has a large line-of-sight velocity dispersion (110\,$\rm km \, s^{-1}$). As shown in Figure~\ref{fig:ftr}, AT1's spectrum is visibly different from the other three thick disc bins: it has a much wider ${\rm H}\upbeta$ absorption line and stronger Fe\,\textsc{i} 5270 lines, indicating it is dominated by a younger population and has a higher [M/H], as estimated by \textsc{ppxf}.

Our explanation for this strange “geometrically-thick chemical-thin disc” is that although IC 2531 appears to be a perfectly edge-on galaxy, its galactic disc is actually distorted. We propose that AT1 contains a warped or twisted, or corrugated thin disc region crossing the line-of-sight, rather than simply a thick disc region, such that light from the warped thin disc dominates in this bin. This scenario is consistent with \citet{2015A&A...582A..18A}’s tilted-ring analysis of the H\,\textsc{i} data for IC 2531, in which they report a warp with a maximum tilt about 4° along the line-of-sight. As IC 2531's angular radius is about 200", this seems to be consistent with a warp that reaches the height of northern thick disc bins (about 14" above the middle plane).

There are some other hints that IC 2531 may not be as flat and symmetric as it looks: (1) In optical images (cf. Figure~\ref{fig:posab}), the east (pos A) end of the galaxy disc appears to be thicker, extends further, and more diffuse than the west (pos B) end. The dust lane cuts across the disc from east to west: on the east end the dust lane appears to twist up to the top (north) of the visible disc, but on the west end the dust lane appears to twist down to the bottom (south) of the visible disc; (2) comparisons of stellar populations show that pos A northern half is most similar to the Milky Way’s low $z$ plane, while other regions are most similar to the Milky Way’s middle and high $z$; (3) as shown in Figure~\ref{fig:m shape} and \ref {fig:scaleheight vs R}, the northern half of pos A is dimmer and have significantly smaller scale height than the southern half over most of the disc; (4) considering the thick disc bins, BT1 and BT2 have the same stellar velocity, but the stellar velocity of AT1 (north) is 20\,$\rm km \, s^{-1}$ lower than AT2 (south); (5) from Figure~\ref{fig:ABvs-vg}, it is clear that the northern and southern thick disc components are exactly opposite in $(v_{star}$-$v_{gas})$: in the northern thick disc bins, the ionised gas lies closer to tangent points; in southern thick disc bins, stars are closer to tangent points. This suggests BT1 may be also affected by the warp or other distortion; (6) the thin disc components in both positions show an asymmetric feature in abundances, whereby the southern half \textbf{in projection} is more alpha-rich than the northern half (cf. Table 2); (7) the distribution of H\,\textsc{i} may be more extended along the line-of-sight on the east side of IC 2531 than the west side, as Figure~\ref{fig:gs2kz} shows that the extent of the H\,\textsc{i} $(p,v)$ diagram covered on the east side is much larger than on the west side, and is much more extended than optical velocities in pos A; (8) the ${\rm H}\upbeta$ emission line peak intensities of southern thick discs are stronger than those of northern thick discs (cf. Table~\ref{tab:avg vl}). However, the ${\rm H}\upgamma$ emission is too weak to be precisely measured in most of our bins.

\begin{figure}
    \centering
    \begin{minipage}{0.99\linewidth}
        \centering
        \includegraphics[width=1\linewidth]{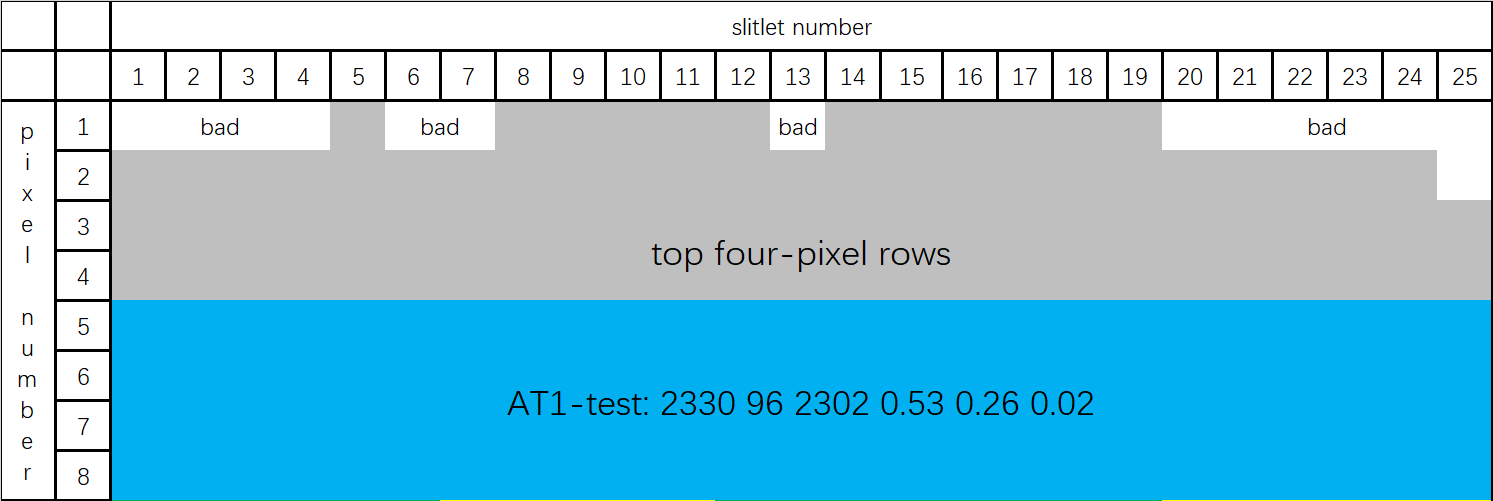}
    \end{minipage}
    \begin{minipage}{0.99\linewidth}
        \centering
        \includegraphics[width=1\linewidth]{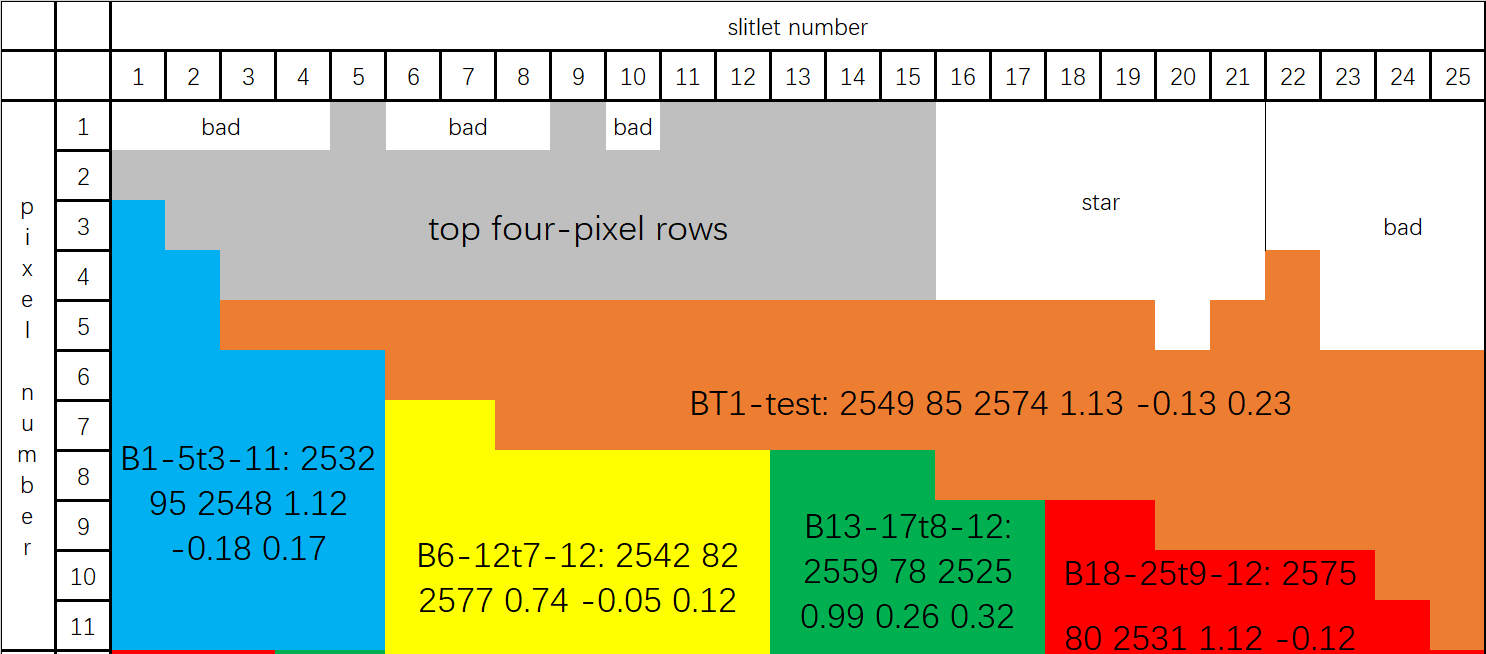}
    \end{minipage}  
    \caption{An illustration of the two test bins AT1-test (top panel) and BT1-test (bottom panel). Grey areas indicate the top four-pixel rows that are removed from AT1 and BT1, cf. Figures~\ref{fig:binac} and~\ref{fig:binbc}.}
  \label{fig:test}
\end{figure}

In each of AT1 and BT1, the top four-pixel rows on their own have inadequate SNR to reliably fit using \textsc{ppxf}; however, we were curious to see their effect upon the estimated stellar kinematics and stellar populations of the northern thick disc bins. We therefore constructed a second set of thick disc test bins in the northern half of the galaxy without the top four-pixel rows included (i.e., only with $y>4$) and repeated our analysis on these bins using \textsc{ppxf}. They are referred as AT1-test and BT1-test in Tables~\ref{tab:MCA} and~\ref{tab:MCB}, and their spatial distributions are illustrated in Figure~\ref{fig:test}. These two test bins are not included in other figures in this paper.

For a subset of stellar parameters, including the top four-pixel rows in bins AT1 and BT1 has a similar effect on both bins. The most significant influence is in stellar velocity dispersion: including these rows increases velocity dispersion by 14 and 18\,$\rm km \, s^{-1}$ respectively, making AT1 and BT1 some of the bins with the highest velocity dispersion. A comparison of BT1 and BT1-test shows that including these rows decreases the [$\alpha$/Fe] by 0.08. This influence is not found in AT1-test, as its [$\alpha$/Fe] already reaches the lower limit of available template [$\alpha$/Fe] values. Although including the top four-pixel rows’ data decreases the SNR in AT1 and BT1 compared to the test bins, it is still adequate for analysis with \textsc{ppxf}. Their effect on age is insignificant compared to uncertainties.

Interestingly, including these rows has different effects on other parameters when comparing AT1 and BT1. Compared to AT1-test, both the stellar and ionised gas velocities in AT1 are closer to the tangent point velocities; meanwhile, in BT1, these velocities are further away from the tangent point velocities in comparison with BT1-test. Likely due to the thick disc’s rotation lag, none of the thick discs’ velocities reach tangent point velocities, except for the ionised gas velocity in AT1. The stellar velocity of AT1 is the closest to tangent point velocities among all the thick disc bins (and exhibits no rotation lag relative to corresponding thin disc bins). This suggests we are likely looking into a more inner part of the galaxy in AT1 than in BT1. The influence upon these rows on metallicity is also different between AT1 and BT1: they increase [M/H] from 0.26 to 0.37 in pos A, but decrease [M/H] from $-0.13$ to $-0.38$ in pos B.

These similarities and differences together suggest that, in pos A, the top four-pixel rows in our observations represent inner parts of the galaxy near the tangent points (i.e., with high radial velocities and a metal-rich stellar population), but in pos B these rows are dominated by outer parts of the galaxy which lie far from the tangent points (i.e., with low radial velocities and a metal-poor stellar population). Although the estimated [M/H] in BT1 can be explained without a warp because it is a geometrical thick disc bin, it may also be affected considering the dramatic increase of stellar velocity dispersions compared to BT1-test, and the difference between northern and southern thick disc bins in Figure~\ref{fig:ABvs-vg}. In this case, the warp doesn't appear to increase its metallicity. We therefore suggest that either the warp or twist in pos A is larger than it is in pos B, or that there is a corrugation which reaches its peak in the middle (small actual radii) of the disc in pos A and raises the outer (large actual radii) parts of the thin disc in pos B. As these distortions are probably triggered by similar mechanisms, we refer to them collectively as warp in this paper.

We note that the stellar velocity dispersions in AT1 (110\,$\rm km \, s^{-1}$) and BT1 (102\,$\rm km \, s^{-1}$) are increased compared to those in AT1-test (96\,$\rm km \, s^{-1}$) and BT1-test (85\,$\rm km \, s^{-1}$), whereas the stellar velocity dispersions of the thin disc bins in pos A and pos B generally range from 70 to 95\,$\rm km \, s^{-1}$. The increase in stellar velocity dispersion is probably due to different orbital speeds or different dispersion components. For a small localised warp, the spread of velocities is due to the integration of light from stars of different orbits with different line-of-sight speeds generating a relatively large velocity dispersion in this bin. Alternatively, if the warp lies a few kpc beyond the major axis of the galaxy along the line-of-sight, the observed velocity dispersion has a significant contribution from the radial component ($\sigma_r$) of the thin disc’s velocity dispersion, which is typically 1.4 times the azimuthal velocity dispersion ($\sigma_{\phi}$, what we normally measure near the tangent points along the line-of-sight). 

\subsubsection{More on warps}

Apart from \citet{2015A&A...582A..18A}, there are other researches suggesting the disc of IC 2531 is distorted as well. 
The H\,\textsc{i} observations reported by~\citet{2004MNRAS.352..768K} revealed evidence for a weak antisymmetric warp along the projected major axis of IC 2531. The warp is larger on the side containing pos A, and is also observed in the optical. The H\,\textsc{i} warp starts at a radius of about 200" in projection, but the position of our proposed warp along the line-of-sight is not observationally determined. 
\citet{2020MNRAS.495.3705N} report the detection of corrugation in the stellar and dust mid-plane of IC 2531, by tracing coordinates of the centre of the vertical dust distribution. They suggest IC 2531’s disc is experiencing a non-axisymmetric perturbation\footnote{We notice there is a scale error of IC 2531 in figure 6 of their paper.}. Although these detected warp and corrugation are perpendicular to our line-of-sight, while our proposed warp/corrugation is along our line-of-sight, similar structures could also exist in other directions.

The origin of stellar and H\,\textsc{i} warps is not well understood. IC 2531's warp might be triggered by tidal interaction with its two companion galaxies, as suggested by~\citet{1998ApJ...504L..23S} for the edge-on galaxy NGC 5907. \citet{2007A&A...466..883V} proposed that the inner flat disc and the outer warped H\,\textsc{i} disc may be diverse components with different origins, such that the inner disc forms earlier with the outer warped H\,\textsc{i} disc forming later from gas infall. If this scenario also applies to the warped stellar disc, then it could explain the young, metal-rich and alpha-poor properties of AT1.

Warped discs in edge-on spiral galaxies are common: \citet{2006NewA...11..293A} found that 236 spirals in their sample of 325 galaxies exhibit optical warps. For edge-on galaxies, the optical warp could be undetected if the warp tilt angle is small or if the maximum deviation from the central plane appears along the line-of-sight. In this case the warp could be mistaken for “thick discs” in the absence of supporting H\,\textsc{i} observations~\citep{2003A&A...405..969G}.

The potential warp we detected in IC 2531 illustrates the difficulty of studying edge-on galaxies, where the underlying structure may not be fully apparent until the analysis is complete. Although an edge-on galaxy may appear perfectly edge-on and flat, this may not be correct, especially as warps are sometimes not evident in structures visible at optical wavelengths. This highlights the benefit of auxiliary H\,\textsc{i} observations of these galaxies.

\section{Summary and discussion}
\label{sec:4}

In this paper, we analyzed integral field spectroscopy data from WiFeS using \textsc{ppxf} fitting and Vazdekis models. Our main conclusions are as follows:

1, IC 2531’s disc above the dust plane generally has stellar populations between the Milky Way’s chemical thin and thick disc. Compared to its geometrical thin disc bins, its geometrical thick disc bins generally have a larger age (>9 Gyr), lower [M/H] ($<-0.25$) but similarly alpha-enhanced [$\alpha$/Fe] (0.15-0.3), one of the thick disc bins (BT2) clearly shows as an alpha-rich ([$\alpha$/Fe] $=0.31$) thick disc. Compared to the Milky Way’s thick disc, it has a similar age, slightly higher [M/H] and lower [$\alpha$/Fe]. There is also a negative radial [M/H] gradient (about $-0.046$ dex/kpc for the mean population) in the thin disc, as in the Milky Way.

2, We found indications of a warped thin disc at a projected location where we expected to find the thick disc, at the northern half of pos A. It has all the characteristics of a thin disc's stellar population: young (about 4 Gyr), metal-rich (0.37), alpha-poor (0.03). However it has a large line-of-sight stellar velocity dispersion (110\,$\rm km \, s^{-1}$) which may be due to an actual increase in velocity dispersion or to projection effects leading to different components of velocity dispersion dominating the kinematics along the line-of-sight: e.g. ($\sigma_r$) rather than ($\sigma_{\phi}$). Pos B may also be slightly affected by this warp. This again illustrates the projection issue in edge-on galaxies and shows the value of checking H\,\textsc{i} observations for potential warps before conducting optical spectroscopy observations, as the warp may not be obvious in the optical.

3, When dust blocks light further along the line-of-sight, it can provide some insight into the radial variation of parameters. Combining our kinematic results with the assumption that dust's optical depth is lower at higher altitudes (so we look deeper into the galaxy), we see pos A's outer half lies on the flat part of IC 2531's rotation curve, while pos A's inner half and pos B mainly lie on the linear part of the rotation curve. Some dust lane bins have low [M/H] and enhanced [$\alpha$/Fe] (especially at $y \approx 17$), like thick discs. This phenomenon indicates that the chemical thick disc of IC 2531 extends in radius beyond the chemical thin disc and much of the galactic dust, which is different from the spatial distribution of the Milky Way’s chemical thick disc. 

The next step is to continue observing more edge-on Milky Way analogues with integral field spectroscopy observations to measure the percentage of similar chemical disc structures, to get a statistically significant conclusion. This will provide further constraints on the formation and evolution histories of such disc galaxies. The ongoing GECKOS survey~\citep{2024IAUS..377...27V} which observes 35 edge-on Milky Way analogues is the first IFS survey driven by this goal.

As discussed throughout this study, projection effects are an inherent problem with edge-on galaxies which complicates the interpretation of galaxies’ kinematics and radial structures. 

It may be possible to study more face-on Milky Way analogues in a similar way, using decomposition techniques to extract the thin and thick disc spectra. This requires a higher spectral resolution and high SNR IFU spectra – see~\citet{2018MNRAS.476.1909A, 2021MNRAS.500.3579A} for related studies of the face-on galaxies NGC 628 and NGC 6946, which shows how the decomposition of spectra can be done using velocity distributions of planetary nebulae and integrated light, in their studies, to separate a younger, kinematically colder stellar component and an older, hotter stellar component in thin discs.

As observed kinematics are dominated by the projection and dynamical effects, some deprojection would be very useful in interpreting the kinematics and population properties. One widely used package is \textsc{dynamite}~\citep{2008MNRAS.385..647V}, a Schwarzschild orbital superposition approach program, which has been applied to the follow up analysis of the Fornax 3D project~\citep{2018A&A...616A.121S} and the SAMI galaxy survey~\citep{2022ApJ...930..153S}.

\section*{Acknowledgements}

We are very grateful to the anonymous referee for many constructive suggestions that improved the presentation and interpretation of our data. We thank Prof. Luis C. Ho and the Carnegie-Irvine Galaxy Survey team for allowing us to use their images of IC 2531, and Prof. Rodrigo Ibata, and Prof. Piet van der Kruit for allowing us to reproduce figures from previously published papers. We are grateful to Dr. Chris Lidman for \textsc{pywifes} instruction and providing reduction scripts. We acknowledge Dr. Marie Martig for the discussions and suggestions of possible scenarios, and we are grateful to Dr. Elisabeta Da Cunha for advice on the choice of spectra models. This research has made use of the NASA/IPAC Extragalactic Database, which is funded by the National Aeronautics and Space Administration and operated by the California Institute of Technology. This research has made use of ESASky, developed by the ESAC Science Data Centre (ESDC) team and maintained alongside other ESA science mission's archives at ESA's European Space Astronomy Centre (ESAC, Madrid, Spain).

%%%%%%%%%%%%%%%%%%%%%%%%%%%%%%%%%%%%%%%%%%%%%%%%%%
\section*{Data Availability}

The WiFeS data is available on request.



\bibliographystyle{mnras}
\bibliography{example} % if your bibtex file is called example.bib

%\begin{thebibliography}{99}
%\end{thebibliography}

%%%%%%%%%%%%%%%%%%%%%%%%%%%%%%%%%%%%%%%%%%%%%%%%%%

%%%%%%%%%%%%%%%%% APPENDICES %%%%%%%%%%%%%%%%%%%%%

\appendix
\section{Figures and tables of bins: detailed information}\
\label{sec:5.1}

\begin{figure*}
    \centering
    \includegraphics[width=0.8\linewidth]{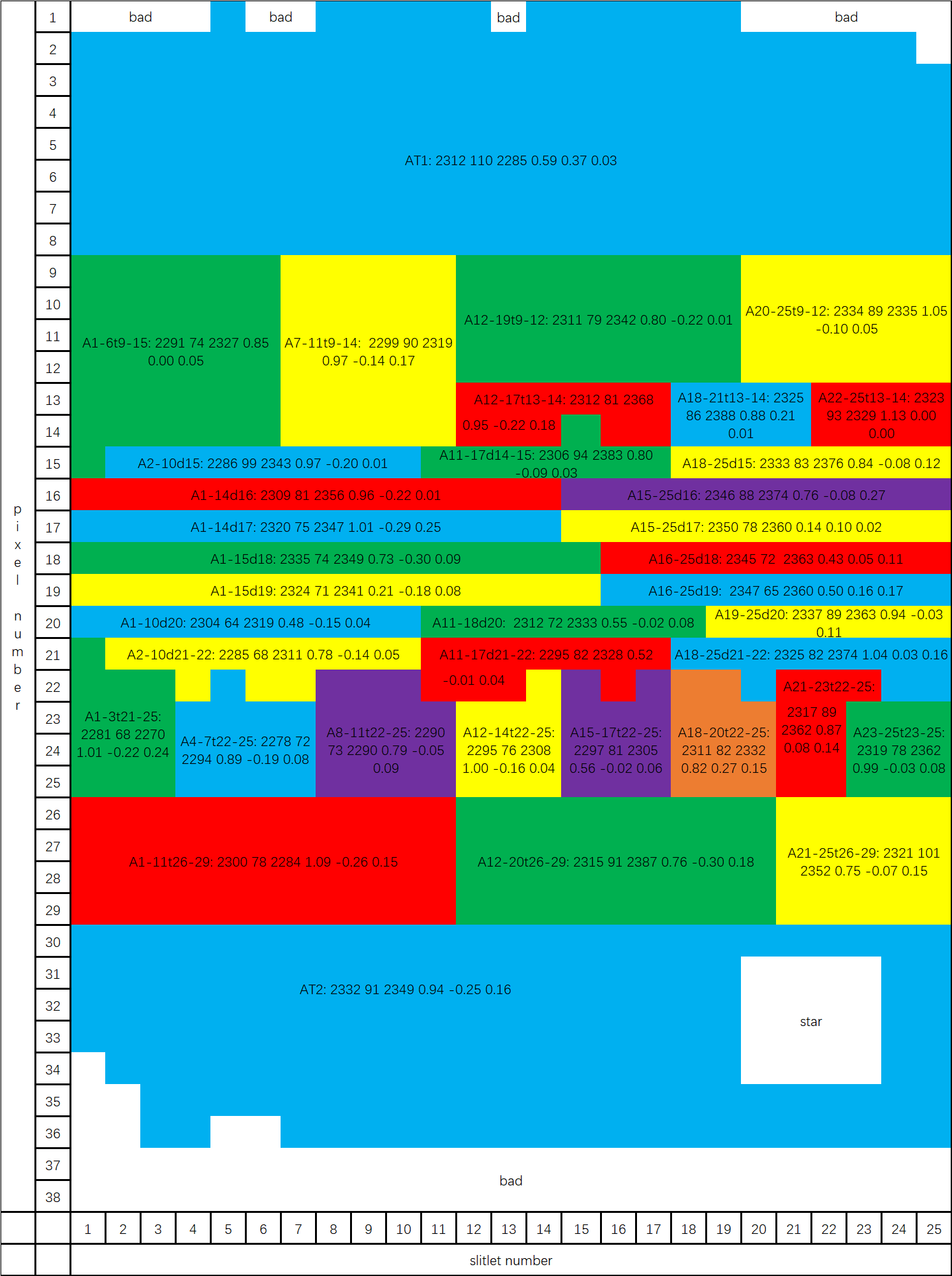}
        \caption{Detailed distribution of bins in pos A and their fitting results. The values in each bin after its name (e.g. A1-11t26-29) represent mean values of (i) stellar velocity (e.g. 2300\,km\,s$^{-1}$), (ii) stellar velocity dispersion (e.g. 78\,km\,s$^{-1}$), (iii) ionised gas velocity (e.g. 2284\,km\,s$^{-1}$), (iv) lg(age) (e.g. 1.09), (v) [M/H] (e.g. $-0.26$), (vi) [$\alpha$/Fe] (e.g. 0.15). The colours in this figure have no meaning and simply represent different bins.}
    \label{fig:binac}
\end{figure*}

\begin{figure*}
    \centering
    \includegraphics[width=0.8\linewidth]{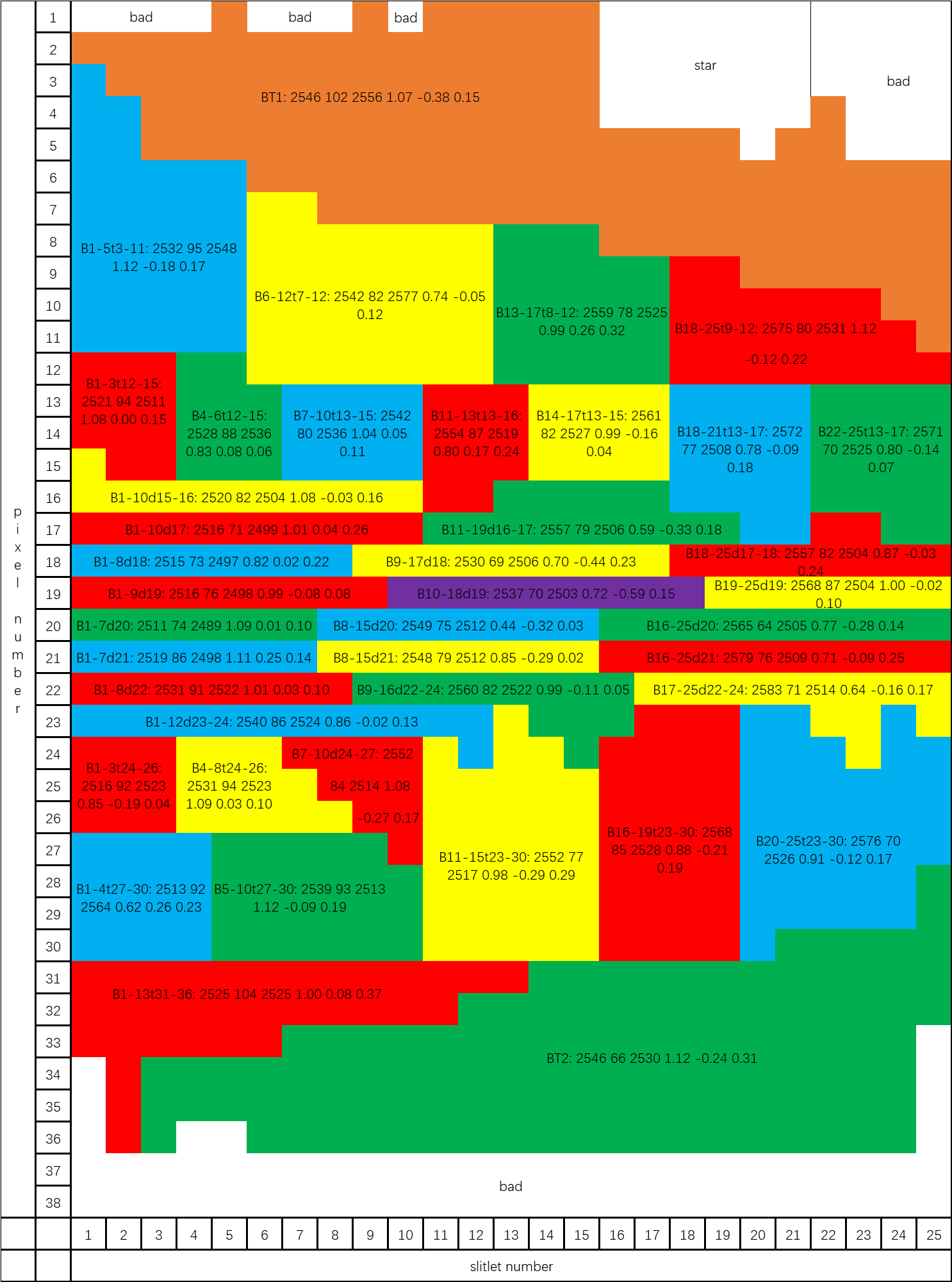}
        \caption{Similar to Figure~\ref{fig:binac} but for pos B.}
    \label{fig:binbc}
\end{figure*}

\begin{table*}
  \centering
  \caption{Detailed information for all the bins in pos A, plus the test bin AT1-test. It includes mean projected radius ("), mean height ("), derived SNR, \textsc{ppxf} fitting kinematics results (km\,s$^{-1}$) and error, \textsc{ppxf} fitting stellar population results and Monte Carlo simulations’ 1$\sigma$ uncertainties, \textsc{ppxf} fitting ${\rm H}\upbeta$ emission line peak intensities (relative flux to units of counts, cf. Figure~\ref{fig:ftr}).
  Bins are arranged in the order of thick disc (northern plus southern), northern thin disc, southern thin disc and the dust lane.}
    \begin{tabular}{|c|c|c|c|c|c|c|c|c|c|c|}
    \hline
          & $\lvert R \rvert$"  & $z$"    & SNR   & star $v$ & star $dv$ & gas $v$ & lg (age) & [M/H] & [$\alpha$/Fe] & ${\rm H}\upbeta$ \\
    \hline
    AT1   & 40.1  & 13.4  & 23    & 2312±4 & 110±4 & 2285±7 & 0.59±0.18 & 0.37±0.08 & 0.03±0.05 & 0.17 \\
   
    AT1-test & 40.0  & 11.7  & 24    & 2330±4 & 96±6  & 2302±5 & 0.53±0.18 & 0.26±0.10 & 0.02±0.05 & 0.29 \\
   
    AT2   & 42.0  & -14.7  & 26    & 2332±4 & 91±4  & 2349±4 & 0.94±0.12 & -0.25±0.10 & 0.16±0.07 & 0.42 \\
   
          &       &       &       &       &       &       &       &       &       &  \\
   
    A1-6t9-15 & 49.5  & 6.2   & 34    & 2291±3 & 74±5  & 2327±3 & 0.85±0.08 & 0.00±0.08 & 0.05±0.05 & 0.37 \\
   
    A7-11t9-14 & 44.0  & 6.7   & 31    & 2299±3 & 90±4  & 2319±6 & 0.97±0.11 & -0.14±0.09 & 0.17±0.07 & 0.23 \\
   
    A12-19t9-12 & 37.5  & 7.7   & 26    & 2311±3 & 79±5  & 2342±5 & 0.80±0.15 & -0.22±0.09 & 0.01±0.02 & 0.28 \\
   
    A20-25t9-12 & 30.5  & 7.7   & 31    & 2334±3 & 89±4  & 2335±7 & 1.05±0.06 & -0.10±0.07 & 0.05±0.04 & 0.09 \\
   
    A12-17t13-14 & 38.6  & 4.8   & 30    & 2312±3 & 81±3  & 2368±3 & 0.95±0.07 & -0.22±0.07 & 0.18±0.05 & 0.36 \\
   
    A18-21t13-14 & 33.5  & 4.7   & 29    & 2325±3 & 86±4  & 2388±2 & 0.88±0.10 & 0.21±0.09 & 0.01±0.04 & 0.47 \\
   
    A22-25t13-14 & 29.5  & 4.7   & 27    & 2323±3 & 93±5  & 2329±8 & 1.13±0.03 & 0.00±0.07 & 0.00±0.02 & 0.06 \\
   
          &       &       &       &       &       &       &       &       &       &  \\
   
    A1-3t21-25 & 51.1  & -5.1  & 35    & 2281±3 & 68±5  & 2270±3 & 1.01±0.05 & -0.22±0.11 & 0.24±0.05 & 0.47 \\
   
    A4-7t22-25 & 47.5  & -5.6  & 37    & 2278±2 & 72±4  & 2294±4 & 0.89±0.08 & -0.19±0.07 & 0.08±0.04 & 0.27 \\
   
    A8-11t22-25 & 43.6  & -5.4  & 35    & 2290±2 & 73±4  & 2290±6 & 0.79±0.07 & -0.05±0.07 & 0.09±0.04 & 0.20 \\
   
    A12-14t22-25 & 39.9  & -5.6  & 31    & 2295±3 & 76±5  & 2308±7 & 1.00±0.07 & -0.16±0.08 & 0.04±0.03 & 0.18 \\
   
    A15-17t22-25 & 37.0  & -5.4  & 36    & 2297±2 & 81±2  & 2305±6 & 0.56±0.12 & -0.02±0.12 & 0.06±0.04 & 0.20 \\
   
    A18-20t22-25 & 34.1  & -5.4  & 32    & 2311±2 & 82±3  & 2332±5 & 0.82±0.07 & 0.27±0.07 & 0.15±0.05 & 0.17 \\
   
    A21-23t22-25 & 31.3  & -5.1  & 31    & 2317±3 & 89±3  & 2362±3 & 0.87±0.09 & 0.08±0.08 & 0.24±0.06 & 0.34 \\
   
    A23-25t23-25 & 29.0  & -5.8  & 29    & 2319±3 & 78±5  & 2362±3 & 0.99±0.07 & -0.03±0.07 & 0.08±0.03 & 0.32 \\
   
    A1-11t26-29 & 47.0  & -9.3  & 26    & 2300±3 & 78±4  & 2284±5 & 1.09±0.06 & -0.26±0.07 & 0.15±0.05 & 0.33 \\
   
    A12-20t26-29 & 37.0  & -9.3  & 28    & 2315±3 & 91±3  & 2387±2 & 0.76±0.09 & -0.30±0.09 & 0.18±0.05 & 0.37 \\
   
    A21-25t26-29 & 30.0  & -9.3  & 29    & 2321±3 & 101±3 & 2352±5 & 0.75±0.09 & -0.07±0.09 & 0.15±0.06 & 0.17 \\
   
          &       &       &       &       &       &       &       &       &       &  \\
   
    A2-10d15 & 47.0  & 3.2   & 22    & 2286±4 & 99±5  & 2343±4 & 0.97±0.11 & -0.20±0.09 & 0.01±0.03 & 0.45 \\
   
    A11-17d14-15 & 38.9  & 3.3   & 28    & 2306±3 & 94±5  & 2383±2 & 0.80±0.11 & -0.09±0.11 & 0.03±0.05 & 0.58 \\
   
    A18-25d15 & 31.5  & 3.2   & 28    & 2333±3 & 83±5  & 2376±3 & 0.84±0.08 & -0.08±0.07 & 0.12±0.05 & 0.30 \\
   
    A1-14d16 & 45.5  & 2.2   & 26    & 2309±3 & 81±5  & 2356±2 & 0.96±0.12 & -0.22±0.10 & 0.01±0.03 & 0.66 \\
   
    A15-25d16 & 33.0  & 2.2   & 25    & 2346±3 & 88±5  & 2374±2 & 0.76±0.15 & -0.08±0.10 & 0.27±0.06 & 0.61 \\
   
    A1-14d17 & 45.5  & 1.2   & 24    & 2320±4 & 75±6  & 2347±2 & 1.01±0.11 & -0.29±0.11 & 0.25±0.08 & 0.78 \\
   
    A15-25d17 & 33.0  & 1.2   & 24    & 2350±4 & 78±7  & 2360±2 & 0.14±0.17 & 0.10±0.05 & 0.02±0.03 & 0.86 \\
   
    A1-15d18 & 45.0  & 0.2   & 23    & 2335±4 & 74±5  & 2349±2 & 0.73±0.17 & -0.30±0.12 & 0.09±0.07 & 0.87 \\
   
    A16-25d18 & 32.5  & 0.2   & 22    & 2345±4 & 72±7  & 2363±2 & 0.43±0.19 & 0.05±0.13 & 0.11±0.07 & 0.90 \\
   
    A1-15d19 & 45.0  & -0.8  & 25    & 2324±4 & 71±5  & 2341±2 & 0.21±0.19 & -0.18±0.16 & 0.08±0.05 & 0.75 \\
   
    A16-25d19 & 32.5  & -0.8  & 24    & 2347±4 & 65±7  & 2360±2 & 0.50±0.19 & 0.16±0.12 & 0.17±0.08 & 0.92 \\
   
    A1-10d20 & 47.5  & -1.8  & 27    & 2304±3 & 64±7  & 2319±2 & 0.48±0.2 & -0.15±0.17 & 0.04±0.05 & 0.89 \\
   
    A11-18d20 & 38.5  & -1.8  & 26    & 2312±3 & 72±6  & 2333±4 & 0.55±0.19 & -0.02±0.11 & 0.08±0.06 & 0.47 \\
   
    A19-25d20 & 31.0  & -1.8  & 22    & 2337±4 & 89±4  & 2363±2 & 0.94±0.16 & -0.03±0.14 & 0.11±0.07 & 0.74 \\
   
    A2-10d21-22 & 47.1  & -3.1  & 32    & 2285±3 & 68±5  & 2311±2 & 0.78±0.13 & -0.14±0.09 & 0.05±0.05 & 0.54 \\
   
    A11-17d21-22 & 39.4  & -3.2  & 37    & 2295±2 & 82±4  & 2328±5 & 0.52±0.13 & -0.01±0.08 & 0.04±0.04 & 0.27 \\
   
    A18-25d21-22 & 31.1  & -3.1  & 36    & 2325±2 & 82±3  & 2374±1 & 1.04±0.06 & 0.03±0.06 & 0.16±0.04 & 0.62 \\
    \hline
    \end{tabular}%
  \label{tab:MCA}%
\end{table*}%

\begin{table*}
  \centering
  \caption{
  Similar to Table~\ref{tab:MCA} but for pos B.}
    \begin{tabular}{|c|c|c|c|c|c|c|c|c|c|c|}
    \hline
          & $\lvert R \rvert$"  & $z$"    & SNR   & star $v$ & star $dv$ & gas $v$ & lg (age) & [M/H] & [$\alpha$/Fe] & ${\rm H}\upbeta$ \\
    
    BT1   & 33.7  & 14.3  & 19    & 2546±5 & 102±4 & 2556±5 & 1.07±0.07 & -0.38±0.06 & 0.15±0.05 & 0.37 \\
   
    BT1-test & 36.9  & 12.7  & 22    & 2549±4 & 85±4  & 2574±5 & 1.13±0.04 & -0.13±0.07 & 0.23±0.04 & 0.32 \\
   
    BT2   & 35.7  & -13.6  & 16    & 2546±5 & 66±7  & 2530±2 & 1.12±0.09 & -0.24±0.12 & 0.31±0.03 & 0.64 \\
   
          &       &       &       &       &       &       &       &       &       &  \\
   
    B1-5t3-11 & 21.9  & 11.8  & 31    & 2532±3 & 95±4  & 2548±6 & 1.12±0.03 & -0.18±0.05 & 0.17±0.03 & 0.20 \\
   
    B6-12t7-12 & 27.9  & 10.1  & 29    & 2542±3 & 82±4  & 2577±5 & 0.74±0.08 & -0.05±0.10 & 0.12±0.04 & 0.17 \\
   
    B13-17t8-12 & 33.9  & 9.8   & 24    & 2559±4 & 78±4  & 2525±5 & 0.99±0.08 & 0.26±0.09 & 0.32±0.04 & 0.28 \\
   
    B18-25t9-12 & 39.8  & 9.0   & 24    & 2575±3 & 80±5  & 2531±3 & 1.12±0.03 & -0.12±0.07 & 0.22±0.04 & 0.38 \\
   
    B1-3t12-15 & 21.1  & 6.6   & 30    & 2521±2 & 94±3  & 2511±3 & 1.08±0.04 & 0.00±0.05 & 0.15±0.03 & 0.33 \\
   
    B4-6t12-15 & 23.9  & 6.3   & 32    & 2528±2 & 88±2  & 2536±3 & 0.83±0.09 & 0.08±0.06 & 0.06±0.05 & 0.35 \\
   
    B7-10t13-15 & 27.5  & 5.9   & 39    & 2542±2 & 80±2  & 2536±3 & 1.04±0.04 & 0.05±0.06 & 0.11±0.04 & 0.32 \\
   
    B11-13t13-16 & 30.9  & 5.6   & 34    & 2554±3 & 87±3  & 2519±3 & 0.80±0.11 & 0.17±0.07 & 0.24±0.06 & 0.42 \\
   
    B14-17t13-15 & 34.5  & 5.9   & 27    & 2561±3 & 82±4  & 2527±2 & 0.99±0.08 & -0.16±0.07 & 0.04±0.04 & 0.48 \\
   
    B18-21t13-17 & 38.6  & 5.2   & 29    & 2572±3 & 77±4  & 2508±0 & 0.78±0.09 & -0.09±0.09 & 0.18±0.05 & 1.72 \\
   
    B22-25t13-17 & 42.6  & 5.2   & 29    & 2571±3 & 70±5  & 2525±1 & 0.80±0.10 & -0.14±0.06 & 0.07±0.05 & 0.68 \\
   
          &       &       &       &       &       &       &       &       &       &  \\
   
    B1-3t24-26 & 21.0  & -5.1  & 37    & 2516±2 & 92±3  & 2523±4 & 0.85±0.07 & -0.19±0.05 & 0.04±0.03 & 0.19 \\
   
    B4-8t24-26 & 24.6  & -5.2  & 44    & 2531±2 & 94±2  & 2523±2 & 1.09±0.03 & 0.03±0.05 & 0.10±0.03 & 0.29 \\
   
    B1-4t27-30 & 21.5  & -8.6  & 26    & 2513±3 & 92±4  & 2564±5 & 0.62±0.10 & 0.26±0.08 & 0.23±0.05 & 0.21 \\
   
    B5-10t27-30 & 26.4  & -8.6  & 28    & 2539±3 & 93±3  & 2513±1 & 1.12±0.04 & -0.09±0.06 & 0.19±0.03 & 0.72 \\
   
    B11-15t23-30 & 32.0  & -7.1  & 38    & 2552±2 & 77±3  & 2517±1 & 0.98±0.05 & -0.29±0.05 & 0.29±0.03 & 0.60 \\
   
    B16-19t23-30 & 36.6  & -6.7  & 34    & 2568±2 & 85±3  & 2528±2 & 0.88±0.09 & -0.21±0.08 & 0.19±0.05 & 0.40 \\
   
    B20-25t23-30 & 41.2  & -6.3  & 33    & 2576±2 & 70±4  & 2526±1 & 0.91±0.07 & -0.12±0.07 & 0.17±0.04 & 0.53 \\
   
    B1-13t31-36 & 24.2  & -12.1  & 21    & 2525±5 & 104±7 & 2520±2 & 1.00±0.08 & 0.08±0.07 & 0.37±0.02 & 0.47 \\
   
          &       &       &       &       &       &       &       &       &       &  \\
   
    B1-10d15-16 & 24.1  & 4.0   & 45    & 2520±2 & 82±3  & 2504±1 & 1.08±0.03 & -0.03±0.04 & 0.16±0.03 & 0.57 \\
   
    B1-10d17 & 24.5  & 2.9   & 42    & 2516±2 & 71±5  & 2499±1 & 1.01±0.05 & 0.04±0.06 & 0.26±0.03 & 0.71 \\
   
    B11-19d16-17 & 34.0  & 3.3   & 32    & 2557±3 & 79±3  & 2506±0 & 0.59±0.10 & -0.33±0.09 & 0.18±0.05 & 1.74 \\
   
    B1-8d18 & 23.5  & 1.9   & 30    & 2515±3 & 73±5  & 2497±1 & 0.82±0.09 & 0.02±0.09 & 0.22±0.05 & 0.83 \\
   
    B9-17d18 & 32.0  & 1.9   & 28    & 2530±3 & 69±6  & 2506±1 & 0.70±0.10 & -0.44±0.11 & 0.23±0.06 & 1.26 \\
   
    B18-25d17-18 & 40.7  & 2.1   & 26    & 2557±4 & 82±6  & 2504±0 & 0.87±0.11 & 0.00±0.08 & 0.24±0.06 & 4.19 \\
   
    B1-9d19 & 24.0  & 0.9   & 26    & 2516±3 & 76±5  & 2498±1 & 0.99±0.08 & -0.08±0.10 & 0.08±0.06 & 0.80 \\
   
    B10-18d19 & 33.0  & 0.9   & 26    & 2537±4 & 70±7  & 2503±1 & 0.72±0.11 & -0.59±0.10 & 0.15±0.06 & 1.53 \\
   
    B19-25d19 & 41.0  & 0.9   & 27    & 2568±4 & 87±7  & 2504±0 & 1.00±0.08 & -0.02±0.11 & 0.10±0.06 & 3.94 \\
   
    B1-7d20 & 23.0  & -0.1  & 26    & 2511±3 & 74±5  & 2489±1 & 1.09±0.06 & 0.01±0.09 & 0.10±0.04 & 0.78 \\
   
    B8-15d20 & 30.5  & -0.1  & 28    & 2549±3 & 75±5  & 2512±2 & 0.44±0.16 & -0.32±0.09 & 0.03±0.04 & 0.78 \\
   
    B16-25d20 & 39.5  & -0.1  & 29    & 2565±3 & 64±6  & 2505±0 & 0.77±0.11 & -0.28±0.15 & 0.14±0.06 & 2.60 \\
   
    B1-7d21 & 23.0  & -1.1  & 32    & 2519±3 & 86±4  & 2498±2 & 1.11±0.03 & 0.25±0.06 & 0.14±0.03 & 0.56 \\
   
    B8-15d21 & 30.5  & -1.1  & 31    & 2548±3 & 79±4  & 2512±2 & 0.85±0.10 & -0.29±0.10 & 0.02±0.03 & 0.53 \\
   
    B16-25d21 & 39.5  & -1.1  & 30    & 2579±3 & 76±4  & 2509±0 & 0.71±0.10 & -0.09±0.10 & 0.25±0.06 & 1.73 \\
   
    B1-8d22 & 23.5  & -2.1  & 38    & 2531±2 & 91±3  & 2522±2 & 1.01±0.04 & 0.03±0.05 & 0.10±0.04 & 0.37 \\
   
    B9-16d22-24 & 32.3  & -2.5  & 41    & 2560±2 & 82±2  & 2522±1 & 0.99±0.05 & -0.11±0.08 & 0.05±0.04 & 0.58 \\
   
    B17-25d22-24 & 40.7  & -2.4  & 34    & 2583±3 & 71±4  & 2514±1 & 0.64±0.09 & -0.16±0.06 & 0.17±0.04 & 1.18 \\
   
    B1-12d23-24 & 25.9  & -3.1  & 51    & 2540±2 & 86±3  & 2524±2 & 0.86±0.06 & -0.02±0.06 & 0.13±0.04 & 0.35 \\
   
    B7-10d24-27 & 28.0  & -5.1  & 36    & 2552±3 & 84±4  & 2514±2 & 1.08±0.04 & -0.27±0.03 & 0.17±0.02 & 0.55 \\
    \hline
    \end{tabular}%
  \label{tab:MCB}%
\end{table*}%

\section{Analysis of light contribution from the bulge in our fields with the CGS image}\
\label{appendix:B}

Pos B is located close to the bulge of IC 2531, and we wish to estimate whether there is significant contamination of the light from pos B by the bulge. The boxy/peanut bulge is a complex triaxial structure which is difficult to model. In addition, the asymmetrical distribution of dust in the disc affects the light distribution of the bulge: see Figure~\ref{fig:posab}. For these reasons, we seek an empirical way to estimate the contribution of the bulge light. 

Using the CGS V-band image, we measured the vertical profiles (parallel to the minor axis) at radial steps of single pixels (0.259") from the center of the bulge out to a projected radius of about 130", and vertically covers $z$ between $\pm$34" of the disc. These profiles were averaged in groups of about 10 pixels for all profiles that were not affected by individual stars.

\begin{figure*}

    \centering
    \begin{minipage}{0.49\linewidth}
        \centering
        \includegraphics[width=1\linewidth]{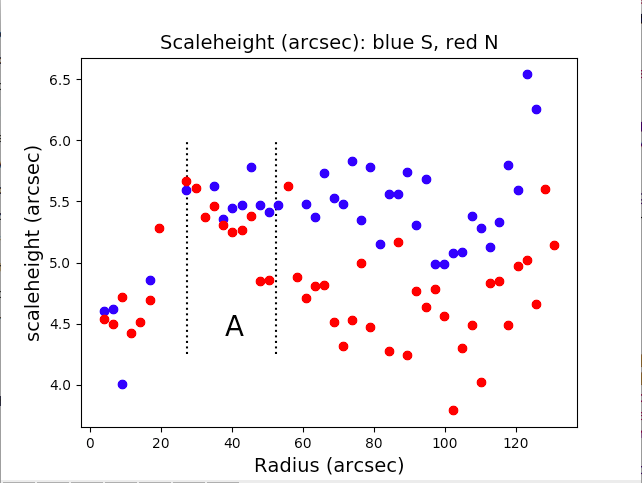}
    \end{minipage}
    \begin{minipage}{0.49\linewidth}
        \centering
        \includegraphics[width=1\linewidth]{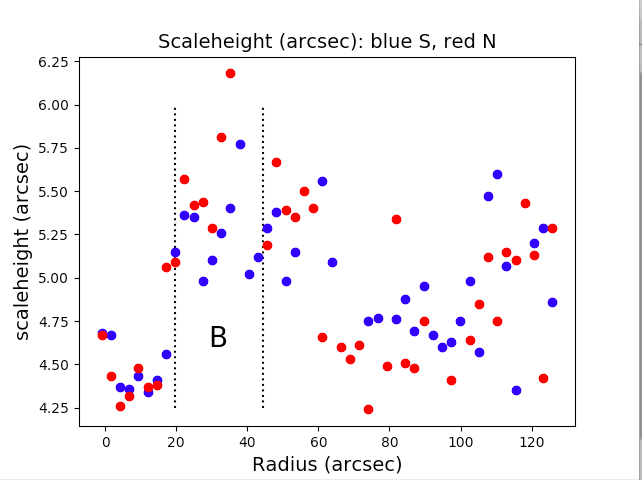}
    \end{minipage}   
    \caption{Measured scale heights of the north (red) and south (blue) sides of IC 2531’s disc using the CGS image, they are derived from the means of about 10 adjacent pixels’ (exclude those affected by stars) vertical M curves (cf. Figure~\ref{fig:m shape}), covering a range of about 130" on each side of the galaxy and vertically covers $z$ between $\pm$34" of the disc. Left panel: IC 2531 pos A (north east) side scale heights vs $R$. Right panel: pos B (south west) side scale heights vs $R$.}
  \label{fig:scaleheight vs R}
\end{figure*}

The profiles showed the usual exponential vertical structure. The scale heights on the north and south sides of the disc were derived and are shown as a function of radius in Figure~\ref{fig:scaleheight vs R}. The locations of the pos A and B are also shown. In the bulge, the scale heights are about 4.5". Moving into the inner disc at a radius around 20" the scale heights abruptly increase to about 5.3", and stay at this level out to about $R\approx50$". Beyond that, the distribution of scale heights becomes broader, as scale heights decrease to about 5" in pos A north, 4" in pos A south and 4.5" in pos B (north and south evolve similarly and have similar values) at $R\approx100$". Finally they all increase with radius again out to the end of the galaxy, probably because the galaxy disc starts to change from thin disc dominated to thick disc dominated at large radius. For the fields on the east side (pos A) of the galaxy, the scale heights on the north side are systematically smaller than on the south.  

We conclude that (1) our pos A and B have scale heights consistent with the inner disc rather than the bulge. Contamination of our fields by the bulge appears to be minimal; (2) there is a strong North-South asymmetry in the vertical structure of the disc on the east side (pos A) of the galaxy.

%%%%%%%%%%%%%%%%%%%%%%%%%%%%%%%%%%%%%%%%%%%%%%%%%%

% Don't change these lines
\bsp	% typesetting comment
\label{lastpage}
\end{document}